\documentclass[letter,11pt]{article}
\pdfoutput=1 

\usepackage{jcappub} 

\usepackage[normalem]{ulem}

\usepackage[utf8]{inputenc}
\usepackage[T1]{fontenc} 
\usepackage{ dsfont }
\usepackage{amsmath,amssymb,calc}
\usepackage{color}
\usepackage{amssymb}
\usepackage{graphicx, epsfig, bm}
\usepackage{hyperref}
\usepackage{soul}
 \usepackage{multicol}
\usepackage{changepage}
\usepackage{subfig}

\usepackage{color}
\usepackage{ifthen}

\def\bdm{\begin{displaymath}}
\def\edm{\end{displaymath}}

\def\barray{\begin{array}}
\def\earray{\end{array}}
\def\be{\begin{equation}}
\def\ee{\end{equation}}
\def\ben{\begin{equation} \nonumber}
\def\een{\end{equation}}
\def\ban{\begin{eqnarray*}}
\def\ean{\end{eqnarray*}}
\def\ba{\begin{eqnarray}}
\def\ea{\end{eqnarray}}
\def\eal{\end{align}}
\def\bal{\begin{align}}
\def\nn{\nonumber}

\def\cH{H}

\def\t{{\rm t}}

\def\({\left(}
\def\){\right)}
\def\[{\left[}
\def\]{\right]}

\def\by{{\bf{y}}}
\def\ba{{a\left(\by\right)}}

\def\bk{{\bf k}}
\def\bq{{\bf q}}

\def\bp{{\bf p}}

\definecolor{gold}{rgb}{1.0, 0.84, 0.0}
\definecolor{maroon}{rgb}{.25,0,0}
\definecolor{darkorange}{rgb}{1.0, 0.55, 0.0}
\definecolor{corn}{rgb}{0.98, 0.93, 0.36}
\definecolor{bronze}{rgb}{0.8, 0.5, 0.2}
\definecolor{darkgreen}{cmyk}{0.85,0.2,1.00,0.2}

\begin{document}

\title{Gravitational waves from fermion production during axion inflation}

\author[a]{Peter Adshead,}
\affiliation[a]{Department of Physics, University of Illinois at Urbana-Champaign, Urbana, Illinois 61801, U.S.A.}
\author[a]{Lauren Pearce,}
\author[b,c]{Marco Peloso,}
\affiliation[b]{Dipartimento di Fisica e Astronomia G. Galilei, Universit\`a degli Studi di Padova, via
Marzolo 8, I-35131, Padova, Italy}
\affiliation[c]{INFN, Sezione di Padova, via Marzolo 8, I-35131 Padova, Italy}
\author[d]{Michael A. Roberts,}
\affiliation[d]{Amherst Center for Fundamental Interactions, Department of Physics, University of Massachusetts, Amherst, MA 01003, U.S.A.}
\author[d]{and Lorenzo Sorbo}


\emailAdd{adshead@illinois.edu}
\emailAdd{lpearce@illinois.edu}
\emailAdd{marco.peloso@pd.infn.it}
\emailAdd{mroberts@umass.edu}
\emailAdd{sorbo@physics.umass.edu}

\abstract{We present analytic results for the gravitational wave power spectrum induced in models where the inflaton is coupled to a fermionic pseudocurrent.  We show that although such a coupling creates helically polarized fermions, the polarized component of the resulting gravitational waves is parametrically suppressed with respect to the non-polarized one.  We also show that the amplitude of the gravitational wave signal associated to this production cannot exceed that generated by the standard mechanism of amplification of vacuum fluctuations. We previously found that this model allows for a regime in which the backreaction of the produced fermions allows for slow-roll inflation even for a steep inflaton potential, and still leads to Gaussian primordial scalar perturbations. The present analysis shows that this regime also results in a gravitational wave signal compatible with the current bounds. }

\maketitle

\section{Introduction}%

There is strong observational evidence supporting the hypothesis that the early universe went through a rapid period of accelerated expansion dubbed inflation~\cite{Guth:1980zm,Linde:2005ht}. As well as resolving the horizon, flatness,  curvature, and gravitino/monopole problems, inflation provides a simple explanation for the observed red-tilted, approximately Gaussian and adiabatic density fluctuations~\cite{Ade:2015xua, Ade:2013ydc}.  The inflationary scenario also generically predicts primordial gravitational waves, which can be measured or constrained through the $B$-mode polarization of the cosmic microwave background. Current measurements restrict the tensor-to-scalar ratio to $r\lesssim0.09$ \cite{Ade:2015tva, Array:2015xqh}, which tightens to $r\lesssim 0.06$ when the  consistency relation $n_{t} = -r/8$, as appropriate for vacuum fluctuations from slow-roll inflation, is imposed. Future experiments are expected to reach $\sigma_r \sim 0.001$ \cite{Abazajian:2016yjj}.  

Slow-roll inflation requires a flat potential; in axion (or natural) inflation this flatness is a natural consequence of an approximate shift symmetry~\cite{Freese:1990rb}.  Axion, or pseudoscalar, models can be motivated within string theory, in which case monodromy~\cite{Silverstein:2008sg, McAllister:2008hb,Kaloper:2008fb,Kaloper:2011jz,DAmico:2017cda} or alignment~\cite{Long:2014dta,Bachlechner:2014hsa,Peloso:2015dsa, Deshpande:2019kjl} effects can give rise to vacuum primordial gravitational waves within the reach of future experiments. 

Due to the approximate shift symmetry, the axion inflaton $\varphi$ must couple derivatively to matter fields.  At mass-dimension five, the possible couplings are 
\begin{equation}
{\Delta \mathcal{L}} = \frac{\varphi}{f} F {\tilde F} + \frac{\partial_\mu \varphi}{f} \, {\bar X} \gamma^\mu \gamma_5 X
\;, 
\label{phi-A-psi}
\end{equation}
to  gauge fields and fermions $X$, respectively. Here $F$ is the usual gauge-field field-strength tensor, $\tilde{F}$ is its dual, and $f$ is a scale known as the axion decay constant.  This coupling of the axion to gauge fields  leads to exponentially large gauge field amplification, with several possible phenomenological consequences (see \cite{Pajer:2013fsa} for a review).  These include steep inflation \cite{Anber:2009ua}, thermalized inflation \cite{Ferreira:2017lnd,Ferreira:2017wlx}, magnetic field production \cite{Prokopec:2001nc, Anber:2006xt,Caprini:2014mja, Fujita:2015iga, Adshead:2016iae,Caprini:2017vnn}, large non-Gaussianity \cite{Barnaby:2010vf, Barnaby:2011vw, Barnaby:2011qe}, chiral gravitational wave production \cite{Sorbo:2011rz,Cook:2013xea, Adshead:2013qp, Adshead:2013nka, Shiraishi:2013kxa, Adshead:2018doq}, instantaneous preheating \cite{Adshead:2015pva, Cuissa:2018oiw}, and the generation of primordial black holes \cite{Linde:2012bt, Bugaev:2013fya,Garcia-Bellido:2016dkw}.

The fermionic coupling has not attracted as much attention, first being studied by one of us in reference~\cite{Adshead:2015kza}, and, more recently, in references \cite{Adshead:2018oaa,Domcke:2018eki,Min:2018rxw}. Due to Pauli blocking, fermions cannot undergo the same exponential amplification as the gauge fields. Furthermore, on the one hand, massless fermions are conformal and therefore cannot be created gravitationally through the expansion of the Universe~\cite{Parker:1968mv}. On the other hand, very heavy fermions decouple, so that, in the absence of the coupling~(\ref{phi-A-psi}), only fermions with mass $m\approx H$ are produced in sizable quantities.  With only one scale in the problem, the Hubble scale, the energy density is $\sim H^4$, which is too small to produce observable effects (with the possible exception of super-heavy dark matter~\cite{Kuzmin:1998kk,Chung:2011ck}). During axion inflation, however, this conclusion does not follow due to the presence of the additional scale $\dot\varphi/f$.    

We have previously studied the regime $\dot \varphi \slash f \gg H$ in reference \cite{Adshead:2018oaa}.  Because fermion modes can be populated up to $k \sim \dot \varphi \slash f$, the energy density can be parametrically larger than $H^4$, as first noticed in reference \cite{Adshead:2015kza}.  In reference  \cite{Adshead:2018oaa} we also identified a regime in which the sourced contribution to the power spectrum dominates the vacuum contribution, yet the non-Gaussianity is beneath current observational bounds.  This behavior is in striking contrast to the analogous effect in systems with strong bosonic particle production. This difference can be understood by noting that phenomenologically interesting results require one to populate a large number of fermion modes and, since the occupation of these modes is restricted by Pauli exclusion, their sum is uncorrelated and becomes increasingly Gaussian by the central limit theorem. Bosonic systems, conversely, allow for large occupation numbers \emph{per mode}, which add coherently  leading to sourced $n$-point functions that are generically related to the two-point function by $\langle \delta\varphi^n\rangle \sim \left( \langle \delta\varphi^2\rangle \right)^{n/2}$ \cite{Barnaby:2011pe}.  As the fermion-axion coupling is increased, eventually one enters a regime of strong backreaction where the evolution of the inflaton zero-mode is controlled by particle production; this is the fermionic analogue of steep inflation studied in reference \cite{Anber:2009ua}. We expect that the above argument that non-Gaussianity remains small still holds in the regime of strong backreaction.

Given the rich phenomenology of the scalar perturbations found in reference \cite{Adshead:2018oaa}, it is  important to characterize how fermions source gravitational waves in the regime where $\dot \varphi \slash f \gg H$.  As was noticed in reference \cite{Adshead:2015kza} and confirmed in reference \cite{Adshead:2018oaa}, the produced fermions have a helicity asymmetry, which was used for leptogenesis in reference \cite{Adshead:2015jza}.  This helicity asymmetry raises the possibility that the spectrum of sourced gravitational waves has a chiral component. Sourced production of gravitational waves in this context was previously studied\footnote{Gravitational wave production by non-chiral fermions has also been studied in reference~\cite{Figueroa:2013vif}.}  in reference~\cite{Anber:2016yqr}. However, the fermion basis used in that work was leading to pathologies as $m\to 0$, whereas in this work we use the basis introduced in reference \cite{Adshead:2018oaa}, in which perturbation theory remains valid as the fermion mass $m$ becomes small. We review this basis while introducing our model in section~\ref{sec:action}.

This paper is organized as follows. In section~\ref{sec:action}, we introduce our theory, and working in the Arnowitt-Deser-Misner (ADM) formalism, we solve the gravitational constraint equations to second order. We then use these solutions to obtain the interaction Lagrangian to fourth order in fluctuations. From this interaction Lagrangian, we obtain one ${\cal O}(\gamma\Psi^2)$ vertex and seven ${\cal O}(\gamma^2\Psi^2)$ vertices. In section~\ref{sec:loop_diagrams}, we use these interactions to compute eight loop diagrams in the in-in formalism. The ${\cal O}(\gamma^2\Psi^2)$ interactions lead to seven one-loop one-vertex diagrams, which we evaluate in section~\ref{sec:quartic_loops}, while the ${\cal O}(\gamma\Psi^2)$ generates a two-vertex loop, which we evaluate in section~\ref{sec:cubic_loop}. We discuss our results in section~\ref{sec:discussion}; we show that the chirally asymmetric contribution is subdominant, and that the total sourced contribution to the tensor-to-scalar ratio is  beneath the vacuum component.  Details of our calculations can be found in the various appendices. We work in natural units where $\hbar = c = 1$, and $M_{\rm Pl} =1/ \sqrt{8\pi G}$ is the reduced Planck mass.

\section{Fermion-graviton interactions during axion inflation}
\label{sec:action}%

The aim of this work is to compute the production of gravitational waves by the fermions. At leading order in perturbation theory, fermions source gravitational wave power at one-loop. At one-loop, diagrams of two topologies are possible.  The first topology---the cubic loop---is a two-vertex diagram that is generated by two cubic order vertices consisting of one gravitational wave and a fermion bilinear. The second topology---the quartic loop---is a one-vertex diagram generated by a quartic-order vertex consisting of two gravitational waves and a fermion bilinear (see figure \ref{fig:diagrams} below). To find the required interactions, we therefore need to expand the full action to quartic order in fluctuations.

\subsection{Starting action}

We consider a theory containing a pseudoscalar inflaton $\varphi$ with a shift-symmetric coupling to a fermion $X$ and minimally coupled to gravity, so that our action, in mostly minus convention, reads
\begin{align}
S = \int d^4 x  \sqrt{-g}\[\frac{M_{\rm Pl}^2}{2}\,R+\frac{g^{\mu\nu}}{2}\partial_\mu\varphi\partial_\nu\varphi - V(\varphi)+ \bar{X} \left(i\gamma^{\mu}D_{\mu} X- m-\frac{1}{f}\partial_{\mu}\varphi \gamma^{\mu}\gamma_5\right)X\].
\end{align}

As discussed in reference \cite{Adshead:2018oaa} (in particular, see  appendix B of that work for details), the use of these fields makes apparent the shift-symmetric nature of the inflaton-fermion interaction, but it obscures the fact that such interaction vanishes as $m\to 0$. It is therefore convenient  to redefine the fermion according to
\begin{align}
X \to \Psi =  e^{i\gamma_5 \frac{ \varphi}{f}}X \;, 
\end{align}
which puts the fermion action in the form
\begin{align}
S_\Psi = \int d^4 x \sqrt{-g}\,\bar\Psi \left[i\gamma^{C}e_{C}{}^\mu\(\partial_{\mu} + \frac{1}{2}\omega_{\mu AB}\Sigma^{AB}\)\Psi- m\,\cos\(\frac{2\varphi}{f}\)+i m\,\sin\(\frac{2\varphi}{f}\)\gamma_5\right]\Psi,
\label{fermion-action1}
\end{align}
where greek letters are spacetime indices $\mu, \nu \in \{0, 1, 2, 3\}$, capital roman letters are 4D Lorentz indices $A, B, C \in \{0, 1, 2, 3\}$, lower case roman letters from the start of the alphabet are spatial Lorentz indices $a,b,c \in \{1,2,3\}$, and finally roman letters from the middle of the alphabet are spatial spacetime indices, $i,j,k \in \{1,2,3\}$.  The generator of local Lorentz transformations is $\Sigma^{AB} = \frac{1}{4}\left[\gamma^A, \gamma^B\right]$, and the spin connection is
$\omega_{\mu}{}^{AB} = e^A{}_{\nu}\nabla_\mu e^{B\nu}$, where $e^A{}_{\nu}$ is the vierbein.

\subsection{The action in ADM form}

Because certain components of the metric are constrained degrees of freedom whose values depend on the fermion bilinears (as well as the other dynamical degrees of freedom), gravitationally-mediated fermion-graviton couplings are generated when these constraints are eliminated from the action. In order to perform this analysis, it is convenient to decompose the metric using the ADM formalism. The key advantage of this formulation is that the constrained degrees of freedom enter the theory algebraically; their equations of motion are algebraic constraints. The metric in ADM form reads
\begin{equation}
ds^{2} = N^{2}d\tau^{2}-{h}_{ij}(dx^{i}+N^{i}d\tau)(dx^{j}+N^{j}d\tau) \;,
\label{line} 
\end{equation}
where $N$ is the lapse and $N^i$ is the shift. For the background metric we choose $N = a$, so that $\tau$ denotes conformal time. Derivatives with respect to $\tau$ are represented with primes. The spatial indices, $i,\,j,\, k,...$ are raised and lowered with $h_{ij}$, so that $N^i\equiv h^{ij}\,N_j$, $h^{ij}\,h_{jk}=\delta^i_k$. Finally, ${\rm det}\left[g\right]=-N^2\,{\rm det}\left[h\right]$.

In these coordinates, the action for the purely bosonic sector of the theory (involving gravity and the inflation) becomes
\begin{align}
S_{\rm B} =& \int d\tau\,d^{3}x\,N\sqrt{h}\left[\frac{M_{\rm Pl}^2}{2}\,\left({}^{(3)}\!R+K^{ij}\,K_{ij}-K^{2}\right)+\frac{\pi_\varphi^{2}}{2\,N^2}-\frac{1}{2}h^{ij}\partial_{i}\varphi\partial_{j}\varphi -V(\varphi)\right] \;, 
\label{S-bose}
\end{align}
where $\pi_\varphi \equiv {\varphi}' - N^{j}\partial_{j}\varphi$ and
\begin{align}
K_{ij}  \equiv & -\frac{1}{2N}\left({h}_{ij}' -
{}^{(3)}\nabla_{i}N_{j}-{}^{(3)}\nabla_{j}N_{i}\right),\quad  K  =  K^{i}_{\;i}\,.
\end{align}
The fermionic action, $S_F$, in these coordinates reads 
\begin{align}\label{fermionaction}
S_F = \int d^4 x \, \mathcal{L}_F,
\end{align}
where (see appendix~\ref{ap:lagrangian})
\begin{align}\nn
\mathcal{L}_F =    a^3 \Bigg\{  &  i\bar\Psi \gamma^{0} \Big[ \partial_{0}+  \(\partial_i N  - N^j \, K_{ij}\)e_b{}^{i}\Sigma^{0b}  +\frac{1}{2}e^{c}{}_{k}\(\partial_0 e_b{}^{k} -\(   N K^{k}{}_{m} - ^{(3)}\!\nabla_m N^k\)e_b{}^{m}\)\eta_{ac}\Sigma^{ab} \Big] \Psi  \\\nn
& +  i\bar\Psi \(\gamma^{a}Ne_{a}{}^k-\gamma^{0}N^k\)\[\partial_{k} - K_{ik} e_b{}^{i}\Sigma^{0b}+\frac{1}{2}\(e^{c}{}_{i}\partial_k e_b{}^{i}  +^{(3)}\!\Gamma^m{}_{ik}e^{c}{}_{m}e_b{}^{i}\)\eta_{ac}\Sigma^{ab} \] \Psi
 \\ & - N m\bar{\Psi}\[ \cos\(\frac{2\varphi}{f}\)+i \sin\(\frac{2\varphi}{f}\)\gamma_5\]\Psi\Bigg\} \, .
\label{fermion-action21} 
\end{align}
In these expressions, $ ^{(3)}\nabla_i$ denotes the three dimensional covariant derivative, $\eta_{ab}$ is the spatial part of the Minkowski metric, $\eta_{AB} = {\rm diag}[1, -1, -1, -1]$, and the spatial vielbeins satisfy $\delta_{ab}e^a \phantom{}_ie^b \phantom{}_j=h_{ij}$. The total action is the sum $S = S_B+S_F$. 

\subsection{Constraints}

When written in terms of the ADM decomposition, one can see that the lapse $N$ and the shift $N_i$ enter in the action, eqs.\ \eqref{S-bose} and \eqref{fermion-action21}, without time derivatives (in the case of the lapse,  spatial derivatives are also missing from the action). This implies that the corresponding Euler-Lagrange equations are constraints. The equation of motion for the lapse is the Hamiltonian constraint
\begin{align} \label{constr1}
0=\frac{\delta S}{\delta N}=&
\frac{M_{\rm Pl}^2}{2}\,{}^{(3)}R - \frac{1}{2}h^{ij}\partial_i\varphi\partial_j\varphi - V - \frac{M_{\rm Pl}^2}{2}(K^{ij}K_{ij}-K^2) -\frac{1}{2N^2}\pi_\varphi^2\nonumber\\
& + \frac{i}{2}e_a{}^i (\bar{\Psi}\,\gamma^a\, \partial_i\Psi - \partial_i\bar{\Psi}\, \gamma^a\, \Psi) - \frac{1}{4}e_a{}^i\,e_{bj}\,{}^{(3)}\nabla_i e_c{}^j\,\epsilon^{abc}\,\bar{\Psi}\gamma^0\gamma^5\Psi\nonumber\\
&- m \,\bar{\Psi}\Big[\mathrm{cos}\left(\frac{2\varphi}{f}\right) + i\, \mathrm{sin}\left(\frac{2\varphi}{f}\right)\gamma^5\Big]\Psi\,,
\end{align}
while the equation of motion for the shift is the momentum constraint
\begin{align} \label{constr2}
0=\frac{\delta S}{\delta N_i}-{}^{(3)}\nabla_j\frac{\delta S}{\delta ({}^{(3)}\nabla_jN_i)}=& \frac{1}{N}\pi_\varphi\partial_i\varphi + \frac{i}{2}(\bar{\Psi}\,\gamma^0\, \partial_i\Psi - \partial_i\bar{\Psi} \,\gamma^0 \,\Psi)+ \frac{1}{4} e^a{}_j\,{}^{(3)}\nabla_i e^{bj}\,\epsilon_{abc}\,\bar{\Psi}\gamma^c\gamma^5\Psi \nonumber\\
&+M_{\rm Pl}^2{}^{(3)}\nabla^j\left(K_{ij}-h_{ij}\,K\right) + \frac{1}{4}{}^{(3)}\nabla^j\Big( e^a{}_i\,e^b{}_j\,\epsilon_{abc}\,\bar{\Psi}\gamma^c\gamma^5\Psi \Big) \;.
\end{align}
In these expressions, ${\epsilon}^{abc}$ is the ``flat'' three-dimensional Levi-Civita tensor, with the convention ${\epsilon}^{123}=+1$. 

We work in the spatially flat gauge where $\det[h_{ij}] = a^6$ and the dynamical scalar fluctuation degrees of freedom are in the fluctuations of the inflaton. We parametrize the tensor perturbations of the metric as\footnote{Repeated lower roman indices are summed with the Kronecker delta: $x_{i} x_{i} \equiv \sum_{i,j = 1}^3 \delta^{ij} x_i x_j $} 
\begin{equation}
h_{ij} = a^2 \left( {\rm e}^{\gamma} \right)_{ij} = a^2 \left[ \delta_{ij} + \gamma_{ij} + \frac{1}{2} \gamma_{im} \gamma_{mj} + \dots \right] \,, 
\label{hij-exp}
\end{equation} 
where $a( \tau )$ is the scale factor. The transverse-traceless nature of the tensor modes, $\delta^{ij} \gamma_{ij} = \gamma_{ij,j} = 0$, implies that $\det\[e^\gamma\]=1$. Similarly, the spatial vielbeins are expanded in terms of the tensor perturbations as
\begin{align}
e^a{}_{i} =  a\, \delta^{ak}e^{\frac{1}{2}\gamma_{ki}} = a\, \delta^{ak}\[1+\frac{1}{2}\gamma_{ki}+\frac{1}{8}\gamma_{kj}\gamma_{ji}\ldots \]. 
\end{align}

Our  goal is to determine the effective cubic $\gamma\bar\Psi\Psi$ and quartic  $\gamma\gamma\bar\Psi\Psi$ component of the Lagrangian, where $\gamma$ schematically denotes the graviton. As is well known, in order to determine the action to $n$-th order in the fluctuations, the solutions to the constraint equations are required at order $n - 2$; terms of order $n$ and $n-1$ simply multiply lower-order constraint equations \cite{Chen:2006nt}. Thus, to obtain the action up to fourth order in the fluctuations, we require solutions for the constraints (the lapse and shift) up to quadratic order. Note that Lorentz invariance means that the fermion fields only begin to contribute to the constraint equations at quadratic order. We therefore solve the above constraints in eqs.\ \eqref{constr1} and \eqref{constr2} perturbatively, to second order in the fluctuations, before plugging them back into the original action. 

To facilitate a perturbative solution, we expand the lapse and shift functions as 
\begin{align}
N &= a\left(1 + \alpha^{(1)} + \alpha^{(2)} + \dots \right) \;, \nonumber \\
N_i &= \partial_i \theta^{(1)} + \partial_i \theta^{(2)} + \dots 
+ \beta_i^{(1)} + \beta_i^{(2)} + \dots \;, 
\label{al-be-th}
\end{align}
where the superscript denotes the order of the expansion, and where $\beta_i^{(1,\,2)}$ are transverse, $\partial_i^{(1,\,2)} \beta_i = 0$. We also expand the inflaton as $\varphi(\tau,\,\vec{x})=\varphi_0(\tau)+\delta\varphi(\tau,\,\vec{x})$, and we  treat the fermions as first order quanties, so that fermion bilinears are of second order.

To zeroth order, the Hamiltonian constraint reduces to the Friedmann equation
\begin{align}
{\cal H}^2=\frac{1}{3\,M_{\rm Pl}^2}\left(\frac{\varphi_0'{}^2}{2}+a^2\,V(\varphi)\right)\,,\qquad {\cal H}\equiv \frac{a'}{a} \;, 
\label{constsol-0}
\end{align}
while the momentum constraint is automatically satisfied. At first order, we obtain~\cite{Maldacena:2002vr}
\begin{align}
\beta^{(1)} = 0 \,,\qquad
\alpha^{(1)} = \dfrac{\varphi_0^\prime }{2\, {\cal H} \,M_{P}^2} \, \delta \varphi\,,\qquad
\Delta \theta^{(1)} = -\dfrac{\varphi_0^{\prime 2} }{2\,  M_{P}^2 {\cal H}^2}  
\left( \dfrac{\mathcal{H}\, \delta \varphi}{ \varphi_0^\prime} \right)^\prime.
\label{constsol-1}
\end{align}
Since we are not interested in the perturbations sourced by fluctuations of the inflaton, we drop these from now on. Ignoring inflaton fluctuations, the second order constraints read
\begin{align}
 \alpha^{(2)}  
&= \Delta^{-1} \left\lbrace \dfrac{1}{8 \mathcal{H}} \partial_j \left[
(\partial_j \gamma_{\ell i}) \gamma^\prime_{i \ell} 
 \right] 
+ 
\dfrac{ia}{4  M_{\rm {Pl}}^2  \mathcal{H}}
 \left[
\bar{\Psi} \gamma^0 \Delta \Psi
- (  \Delta \bar{\Psi}) \gamma^0 \Psi
 \right] \right\rbrace, \nonumber \\
\beta_j^{(2)}
&= \Delta^{-1} \left\lbrace
\dfrac{1}{2} \Delta^{-1}  \partial_j \partial_k \left[
(\partial_k \gamma_{\ell i}) \gamma^\prime_{i \ell}
 \right] 
 - \dfrac{1}{2} \left[
(\partial_i \gamma^\prime_{jk}) \gamma_{ki}
+(\partial_j \gamma_{\ell i}) \gamma^\prime_{i \ell}
-(\partial_i \gamma_{jk}) \gamma^\prime_{ki}
\right]
\right. \nonumber \\
& \qquad \left.
+ 
\dfrac{ia}{M_{\rm {Pl}}^2 }
\partial_j \Delta^{-1} \left[
\bar{\Psi} \gamma^0 \Delta \Psi
- (\Delta \bar{\Psi}) \gamma^0 \Psi
 \right] \right. \nonumber \\
& \qquad \left.
- \dfrac{a}{M_{\rm {Pl}}^2} \left[i \left( \bar{\Psi} \gamma^0 \partial_j \Psi - (\partial_j \bar{\Psi}) \gamma^0 \Psi \right)
- \dfrac{1}{2}  \epsilon_{ijk } \partial_i (\bar{\Psi} \gamma^k \gamma^5 \Psi) 
\right]
\right\rbrace ,\nonumber \\
\theta^{(2)}
&= \Delta^{-1} \left\lbrace - 
 \dfrac{1}{16 \mathcal{H}} \left[ \gamma^\prime_{ij} \gamma^\prime_{ij}
+  \left( \partial_j \gamma_{kq}  \right)\partial_j \gamma_{qk}  \right] 
-  \dfrac{ia}{4  M_{\rm {Pl}}^2 \mathcal H}   
\left( \bar{\Psi} \gamma^0 \partial_0 \Psi - (\partial_0 \bar{\Psi}) \gamma^0  \Psi \right) \right. \nonumber \\
& \qquad \left.
-  \dfrac{a^2}{M_{\rm {Pl}}^2 \mathcal H} V( \varphi_0) 
\Delta^{-1} \left\lbrace \dfrac{1}{8 \mathcal{H}} \partial_j \left[
(\partial_j \gamma_{\ell i}) \gamma^\prime_{i \ell}
 \right] 
+ 
\dfrac{ia }{4  M_{\rm {Pl}}^2  \mathcal{H}}
 \left[
\bar{\Psi} \gamma^0 \Delta \Psi
- (  \Delta \bar{\Psi}) \gamma^0 \Psi
 \right] \right\rbrace   \right\rbrace,
\label{constsol-2}
\end{align}
where $\Delta = \partial_i\partial_i$ is the spatial Laplacian, $\Delta^{-1}$ is its inverse, and we note that $\epsilon_{123} = -1$.
We have used the linear equation of motion for the fermion to simplify the solution for $\theta^{(2)}$; the details of this calculation are given in appendix \ref{ap:lagrangian}. 

\subsection{Explicit form of the fermion action, and fermion-GW interactions}%

We insert the solutions to the constraint equations for $N$ and $N^i$ to second order (eqs.\ (\ref{constsol-0}), (\ref{constsol-1}), and (\ref{constsol-2})) into the action, eq.\ \eqref{S-bose} + (\ref{fermionaction}), and then expand order by order in the fluctuations.  This gives the quadratic action $S^{(2)} =S^{(2)}_\gamma + S_F^{(2)}$ for the free gravitons and fermions, the cubic action   $S_F^{(3)}$ describing the ${\cal O } \left( \Psi^2 \gamma \right)$ interactions, and the  quartic action  
 $S_F^{(4)}$ describing the ${\cal O } \left( \Psi^2 \gamma^2 \right)$ interactions. 
 
The quadratic action for the gravitons reads 
\begin{align}
S_\gamma^{(2)} = \frac{M_{\rm Pl}^2}{8}\int d^4 x \, a^2\[\gamma_{ij}'\gamma_{ij}' - \partial_k \gamma_{ij} \partial_k \gamma_{ij}\] \;, 
\end{align}
while the quadratic action for the fermions is
\begin{align} 
S^{(2)}_F =  \int d^4 x   \left[  i\bar\psi \(\gamma^{0}\partial_{0}+\gamma^{a} \partial_{a}\)\psi - m a\, \cos\(\frac{2\varphi}{f}\)\bar{\psi}\psi + i ma \, \sin\(\frac{2\varphi}{f}\)\bar{\psi}\gamma_5\psi \right] \,, 
\label{S2F}
\end{align}
where we have rescaled the fermion field according to ${\psi} \equiv a^{3/2}\Psi$. 

At cubic order we find 
\begin{align} 
S^{(3)}_F =   -\frac{i}{2}  \int d^4 x  \:  \gamma_{ij} \, \bar\psi \, \gamma^{i} \, \partial_j \, \psi 
\equiv   \int d^4 x  \, {\cal L}^{(3)}  \, , 
\label{S3F}
\end{align}
and some straightforward, but lengthy, algebra leads to the quartic order action 
\begin{align}
S^{(4)}_F = & \int d^4 x \Bigg\{  \frac{i}{16} \gamma_{ab}\,\gamma_{bk}\(\bar\psi  \gamma^{a} \partial_{k}\psi - \partial_{k}\bar\psi  \gamma^{a}\psi \){+}\frac{1}{16}\gamma'_{aj}\gamma_{jb} \epsilon_{abc}\bar\psi \gamma^c \gamma^5\psi {+}\frac{1}{16}\epsilon_{abc}\gamma_{kc}\(\partial_{a}{\gamma_{kb}}\)\bar{\psi}\gamma^0 \gamma^5\psi \nonumber\\ 
&  + \frac{i}{4}\(1-\frac{V }{4 \,H^2\,M_{\rm Pl}^2}\)\Delta^{-2}\partial_{m}\left(\partial_{m}\gamma_{kn}\,\gamma'_{kn}\right) \(\bar\psi \gamma^{0}\Delta\psi-(\Delta \bar\psi)\, \gamma^{0}\psi\) \nonumber\\ 
& -\frac{\Delta^{-1}}{8} (\gamma'_{jk}\,\partial_{j}\gamma_{ik}-\gamma_{jk}\,\partial_{j}\gamma'_{ik}-\gamma'_{kj}\,\partial_{i}\gamma_{kj}) \left[\epsilon_{aic}\partial_a(\bar\psi  \gamma^c \gamma^5\psi) +2\, i \(\bar\psi \gamma^{0}\partial_{i}\psi-\partial_{i}\bar\psi \gamma^{0}\psi\) \right] \nonumber\\ 
&  -\frac{ i }{32\,aH}\left(\partial_i\gamma_{jk}\,\partial_i\gamma_{jk}+\gamma_{ij}'\gamma_{ij}'\right) \Delta^{-1} \(\bar\psi \gamma^{0}\,\Delta\psi-\Delta\bar\psi \gamma^{0}\psi\) \nonumber\\ 
& -\frac{i}{16\,aH}\(\bar\psi \gamma^{0}\psi' -\bar\psi'\, \gamma^{0} \psi\)\,  \Delta^{-1} \partial_{i} \left(\partial_{i}\gamma_{kj}\,\gamma'_{kj}\right) \Bigg\} 
\equiv   \int d^4 x  \, {\cal L}^{(4)}  \equiv \sum_{i=1}^7 \,   \int d^4 x  \, {\cal L}_i^{(4)}  \,, 
\label{S4F}
\end{align}	
where $H = \dot{a} \slash a$ is the Hubble parameter and $\epsilon_{123} = -1$.  ${\cal L}_1^{(4)}  ,\, {\cal L}_2^{(4)}  ,\, {\cal L}_3^{(4)} $ refer to the three terms in the first line of eq.\ 
(\ref{S4F}), while the remaining ${\cal L}_i^{(4)}$ refer to the other four lines (one term per line). Note that the interactions ${\cal L}_4^{(4)}  ,\, {\cal L}_5^{(4)} ,\, {\cal L}_6^{(4)}  ,\, {\cal L}_7^{(4)} $ arise from integrating out the non-dynamical constraints (the second order parts of the lapse and shift).  From the cubic and quartic Lagrangian densities we find the interaction Hamiltonian densities  
\begin{eqnarray} 
H_{\rm int}^{(3)} \left( \tau \right) &=&  - \int d^3 x \, {\cal L}^{(3)} \;\;, \nonumber\\  
H_{{\rm int},i}^{(4)}  \left( \tau \right) &=& - \int d^3 x \, {\cal L}_i^{(4)} \;\;,\;\; \left( i = 1 ,\, \dots 7 \right) \;\;. 
\label{Hint} 
\end{eqnarray}

To proceed, we expand the tensors in Fourier space as 
\begin{align}
\gamma_{ij}({\bf x}, \,\tau) = \sum_\lambda \int \frac{d^3 k}{(2\pi)^{3/2}} \,\gamma^\lambda_{\bf k}(\tau) \, \Pi_{ij}^{\lambda} \left( {\bf k} \right) \, e^{i{\bf k} \cdot{\bf x}},\;
 \gamma^{\lambda}_{\bf k}(\tau) = \frac{\sqrt{2}}{a(\tau)\, M_{\rm Pl}} \, \t^{\lambda}_{\bf k}(\tau), 
\label{gamma-deco1}
\end{align}
where the field $\t^{\lambda}_{\bf k}$ is canonically normalized. The sum is over the right-handed $(\lambda = + 1)$ and left-handed  $(\lambda = - 1)$ tensor polarizations, with the polarization tensors satisfying 
\begin{align}
\Pi_{ij}^{\lambda} \left( {\bf k} \right)^* =  \Pi_{ij}^{-\lambda} \left( {\bf k} \right) = \Pi_{ij}^{\lambda} \left( -{\bf k} \right) ,\;\;\; 
\Pi_{ij}^{\lambda} \left( {\bf k} \right)\Pi_{ij}^{\lambda'} \left( {\bf k} \right) =  2\delta_{\lambda, -\lambda'} ,\;\;\; 
\epsilon_{abc}{\bf k}_b \Pi_{cd}^{\lambda} \left( {\bf k} \right)  = i \lambda k \Pi_{ad}^{\lambda} \left( {\bf k} \right). 
\end{align}
We also Fourier transform the fermions according to 
\begin{align}
\psi({\bf x}, \tau) = & \int \frac{d^3 k}{(2\pi)^{3/2}} \; \psi_{\bf k} (\tau)\; e^{i{\bf k}\cdot{\bf x}} \,. 
\label{psi-deco1}
\end{align}
In terms of the  fields $\t^{\lambda}_{\bf k}$ and $\psi_{\bf k}$, the quadratic action is 
\begin{align}\nn
S^{(2)}
= &  \int dt \Bigg[ \int d^3 k\Big( i\bar{{\psi}}_{{\bf k}} \(\gamma^{0}\partial_{0}+i\gamma^{a} k_{a}\)\psi_{{\bf k}} -  ma \cos\(\frac{2\varphi}{f}\)\bar{\psi}_{{\bf k}}\psi_{{\bf k}} +i ma \sin\(\frac{2\varphi}{f}\)\bar{\psi}_{{\bf k}}\gamma_5\psi_{{\bf k}}\Big)\\ 
& \quad\quad +\sum_{\lambda} \frac{1}{2} \int d^3 k\left(\partial_{0}\t_{-\bf k}^\lambda \partial_{0}\t_{\bf k}^{\lambda} -\(k^2-\frac{a''}{a}\)\t^\lambda_{-\bf k} \t^{\lambda}_{\bf k}
\right)\Bigg] \;, \label{eqn:actionquad}
\end{align}
and we note that the kinetic terms are canonically normalized.

Inserting eqs.\ \eqref{gamma-deco1} and \eqref{psi-deco1} in the interaction Hamiltonians eq.\ (\ref{Hint}), we obtain the Fourier space Hamiltonian densities we report in appendix \ref{ap:Hint}. 

\section{Fermion contributions to the tensor power spectrum}
\label{sec:loop_diagrams}%

In this section, making use of the interaction Hamiltonians derived in the previous section, we compute the fermion contribution to the gravitational wave two-point correlation function.  After quantizing the free theory, we introduce the in-in formalism and compute the cubic and quartic loops generated by the interactions derived in section \ref{sec:action}. Finally, we end this section by showing how simple scaling arguments concur with our results.

\subsection{Quantization}%
\label{sec:mode_functions}

We canonically quantize the theory by expanding the fields into modes  
\begin{eqnarray}
\hat{\t}^\lambda_{{\bf k}}\left( \tau \right) = \t_k^\lambda\left( \tau \right)a^{\lambda}_{{\bf k}}+ \t_k^{\lambda, *}\left( \tau \right)a^{\lambda, \dagger}_{-{\bf k}} \;\;\;,\;\;\; 
\psi_{\bf k}\left( \tau \right) = \sum_{r = \pm}\(U^r_{\bf k}\left( \tau \right)b^r_{\bf k} + V^r_{-{\bf k}}\left( \tau \right)c^{r,\dagger}_{-{\bf k}}\) \;\;\;,
\end{eqnarray} 
where the creation-annihilation operators for the tensor modes satisfy the commutation relations
\begin{align}
\left[ a^{\lambda}_{{\bf k}} ,\, a^{\lambda'}_{{\bf k}'}{}^\dagger \right] = \delta_{\lambda\lambda'} \, \delta \left( {\bf k} - {\bf k}' \right),
\end{align} 
and the fermionic operators satisfy anti-commutation relations 
\begin{align}
\{b^{r}_{{\bf k}} ,\, b^{r'}_{{\bf k}'}{}^\dagger\}  = \{c^{r}_{{\bf k}} ,\, c^{r'}_{{\bf k}'}{}^\dagger\}  = \delta_{rr'} \, \delta \left( {\bf k}-{\bf k}' \right).
\end{align}
The mode functions $\t_k^\lambda\left( \tau \right)$, and the spinors $U^r_{\bf k}\left( \tau \right)$ and $V^r_{-{\bf k}}\left( \tau \right)$ are solutions of the Euler-Lagrange equations of motion that follow from the action in eq.\ \eqref{eqn:actionquad}. We further decompose the 4-component fermionic spinors into helicity states 
\begin{eqnarray}
&& U^r_{\bf k}\left( \tau \right)= \frac{1}{\sqrt{2}}\(\begin{matrix} u^r_k\left( \tau \right)\, \chi_r({\bf k}) \\ r v^r_k\left( \tau \right)\,  \chi_r \left( {\bf k} \right) \end{matrix}\) \;\;,\;\; 
V^r_{\bf k} \left( \tau \right) = C \, {\bar U}^r_{\bf k} \left(\tau \right)^T  \;, 
\end{eqnarray}
where $C = i\gamma^0 \gamma^2$ is the charge-conjugation operator, and the spinors $\chi_r \left( {\bk } \right)$ are explicitly given by
\begin{eqnarray}
&& \chi_r \left( {\bk } \right) \equiv \frac{k + r \boldsymbol{\sigma} \cdot {\bf k}}{\sqrt{2 k \left( k + k_z \right)}} {\bar \chi}_r \;\;,\;\; {\bar \chi}_+ = \left( \begin{array}{c} 1 \\ 0 \end{array} \right) \;\;,\;\; {\bar \chi}_- = \left( \begin{array}{c} 0 \\ 1 \end{array} \right) \;,
\end{eqnarray}
where $k_z$ is the $z-$component of ${\bf k}$, and $\sigma_i$ are the Pauli matrices. Note that $\chi_r \left( {\bk } \right)$ are helicity eigenspinors which satisfy ${\bf k}\cdot\boldsymbol{\sigma}\chi_r \left( {\bk } \right) = r k \chi_r \left( {\bk } \right) $. We use the Dirac representation for the $\gamma$ matrices,\footnote{In these expressions, $ \mathds{1}$ denotes the $2 \times 2$ identity matrix.} 
\begin{equation}
\gamma^0 = \left( \begin{array}{cc} \mathds{1} & 0 \\ 0 & -  \mathds{1} \end{array} \right) \;\;,\;\;  
\gamma^i = \left( \begin{array}{cc} 0 & \sigma^i \\ - \sigma^i & 0 \end{array} \right) \;\;,\;\; 
\gamma^5 = \left( \begin{array}{cc} 0 &  \mathds{1} \\   \mathds{1} & 0 \end{array} \right) \;. 
\end{equation} 
To obtain solutions to the classical mode equations, we approximate the background inflationary spacetime as de Sitter space and take the evolution of the inflaton to be rolling at a constant speed in cosmic time.  This implies $\varphi_0(\tau)/f= \varphi_0^{\rm {in}}/f- 2 \xi \log \left( x / x_{\rm in} \right)$, with $x \equiv - k \tau$ and $x_{\rm in}\equiv -k\tau_{\rm in}$, where $\tau_{\rm in}$ is some reference time. With these approximations, the mode functions for the fermion field are given by
\begin{align}\label{eq:def_uv}
{u}^r( x )= & \frac{1}{\sqrt{2 x}} \left[ {\rm e}^{i r \varphi_0/f} \, s^r \left( x \right) +  {\rm e}^{-i r\varphi_0/f} \, d^r \left( x \right) \right], \,\nn\\
{v}^r( x ) = & \frac{1}{\sqrt{2 x}} \left[ {\rm e}^{i r \varphi_0/f} \, s^r \left( x \right) -  {\rm e}^{-i r \varphi_0/f} \, d^r \left( x \right) \right] \,, 
\end{align} 
which satisfy the normalization condition $|u^r|^2+|v^r|^2=2$, with  \cite{Adshead:2015kza, Adshead:2018oaa}
\begin{align}\label{eq:def_sd}
s^r \left( x \right)=   {\rm e}^{-\pi r \xi} \, W_{\frac{1}{2}+2 i r \xi ,\, i \sqrt{\mu^2+4 \xi^2}}(- 2 i x)\,,\quad
 d^r \left( x \right)= - i  \, \mu \, {\rm e}^{-\pi r \xi} \, {W}_{-\frac{1}{2}+2 i r \xi ,\, i \sqrt{\mu^2+4 \xi^2}}( - 2 i x ) \,, 
\end{align}
where $W_{\mu,\,\lambda}(z)$ denotes the Whittaker W-function and
\begin{eqnarray} 
\mu\equiv \frac{m}{H}\,,\quad \xi\equiv \frac{\dot\varphi_0}{2fH}\,. 
\end{eqnarray} 
In the same approximation, the  tensor mode functions read
\begin{align}
 \t^{\lambda}_{\bf k}(\tau)=\frac{1}{\sqrt{2k}}\left(1-\frac{i}{k\tau}\right)\,e^{-ik\tau}\,.
\end{align}
In both cases, the integration constants have been chosen so that the solutions match onto the appropriate Bunch-Davies vacuum solution at early times, $x\to \infty$.

\begin{figure}
\centering
\includegraphics[width=0.60\textwidth]{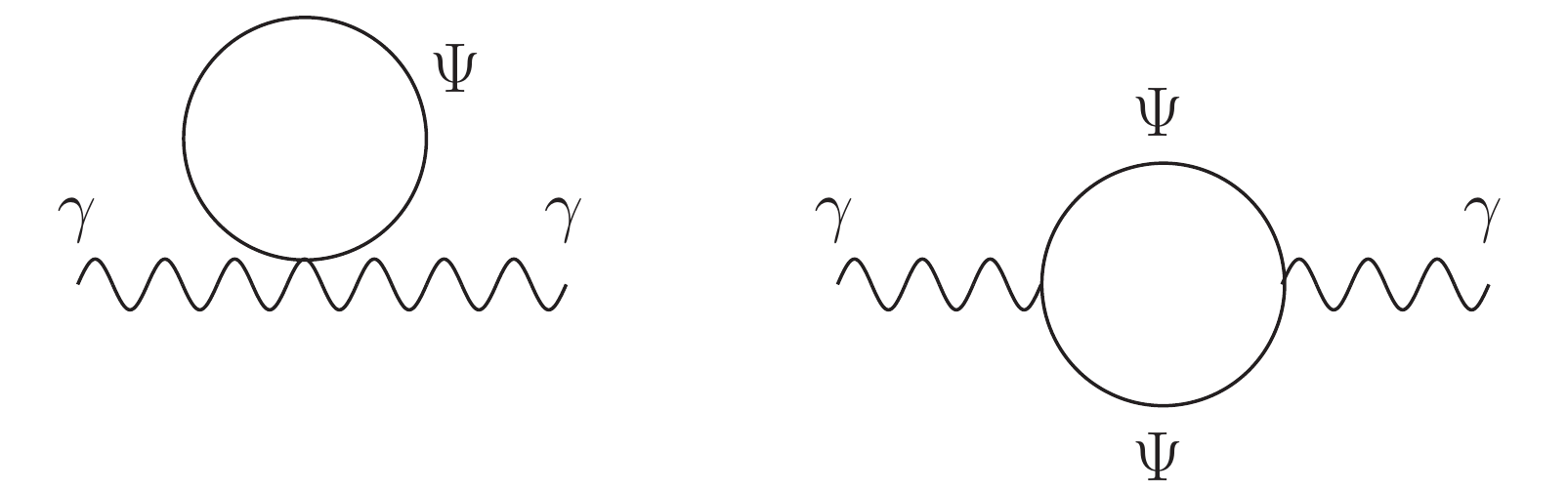}
\caption{The two diagrams that contribute at leading order to the two-point function of the graviton $\gamma$.} 
\label{fig:diagrams}
\end{figure}

\subsection{Fermion loop-corrections to the gravitational wave power spectrum}%
\label{sec:pspectrum}

The interaction Hamiltonians derived above allow us to compute the leading order contributions from the produced fermions to the two-point function of the graviton. These are computed using the in-in formalism, where the correlation function of an operator  ${\hat O}_1  \dots  {\hat O}_n\left( \tau \right)$ at time $\tau$ is given by
\begin{align}
\left\langle {\hat O}_1  \dots  {\hat O}_n\left( \tau \right) \right\rangle =& \sum_{N=0}^\infty 
\left( - i \right)^N \int^\tau d \tau_1 \ldots \int^{\tau_{N-1}} d \tau_N \nonumber\\
&\times\left\langle \left[ \left[ \ldots \left[  {\hat O}_1^{(0)} \dots  {\hat O}_n^{(0)} \left( \tau \right) ,\, H_{\rm int} \left( \tau_1  \right) \right] , \dots \right] ,\, H_{\rm int} \left( \tau_N \right) \right] \right\rangle \,. 
\end{align} 
The interactions in section \ref{sec:action} result in two classes of diagrams: there are seven {\em quartic loop} diagrams, illustrated in the left panel of figure~\ref{fig:diagrams}, one for each of the seven vertices generated by the quartic action~(\ref{S4F}), and one {\em cubic loop} diagram with two vertices generated by the cubic action~(\ref{S3F}), illustrated in the right panel of figure~\ref{fig:diagrams}. We discuss these diagrams in the next two subsections.

\subsubsection{Quartic loops}%
\label{sec:quartic_loops}

We begin with the left diagram of figure \ref{fig:diagrams}. The seven terms in the quartic action lead to seven quartic contributions to the graviton spectrum of the form
\begin{align}\label{eq:quartic_inin}
\left\langle\gamma_{\bp_1}^{\lambda_1}(\tau)\,\gamma_{\bp_2}^{\lambda_2}(\tau)\right\rangle^{(4)}_i=-\frac{2i}{M_{\rm Pl}^2\,a(\tau)^2}\int^\tau d\tau_1\left\langle\left[\t_{\bp_1}^{\lambda_1}(\tau)\,\t_{\bp_2}^{\lambda_2}(\tau),\,H^{(4)}_{{\rm int}, \,i}(\tau_1)\right]\right\rangle\,,\qquad i=1,\,\ldots, 7\,,
\end{align} 
where the interaction Hamiltonians $H^{(4)}_{{\rm int}, i}$ are given in eq.~(\ref{eq:quartic_vertices}). Remarkably, all these diagrams can be computed exactly. The details of the calculation, as well as the exact results, are presented in appendix~\ref{ap:quartic}. Here we summarize the main issues one encounters when performing this calculation. At the end of this section we  present the expression of the sum of the quartic loops in the limit $\mu\ll 1\ll \xi$.

First, several of the terms in the interaction Hamiltonian contain the nonlocal operator $\Delta^{-1}$, the inverse of the Laplacian. When evaluating eq.~(\ref{eq:quartic_inin}) one often encounters the expectation value of quantities evaluated at vanishing momentum that, when acted upon by $\Delta^{-1}$, lead to a undetermined ``$0/0$'' that needs to be regularized. To deal with this limit we follow the prescription given in~\cite{Seery:2007we}: these undetermined quantities are schematically  given by
\begin{align}\label{eq:deltaminus1}
\frac{1}{|\bq_1-\bq_2|^2}f(\bq_1-\bq_2,\,\bp_1)\delta(\bq_1-\bq_2)\,,\qquad f(0,\,\bp_1)=0\,,
\end{align}
where $\bp_1$ is an external momentum. We regularize eq.~(\ref{eq:deltaminus1}) by setting $\bq_1=\bq_2+\boldsymbol{\epsilon}$, where we eventually send $\boldsymbol\epsilon\to 0$. Since $f$ is a scalar, it depends only on $\boldsymbol{\epsilon}\cdot\bp_1$, $p_1^2$, and ${\epsilon}^2$. We then impose that $\boldsymbol{\epsilon}$ approaches zero in a direction that is orthogonal to $\bp_1$, so that $\boldsymbol{\epsilon}\cdot\bp_1={\cal O}({\epsilon}^2)$. With this convention, all the operators containing $\Delta^{-1}$ give finite and unambiguous results.

Secondly, many integrals contributing to the graviton two-point function are divergent in the ultraviolet. We deal with these divergences as we did in~\cite{Adshead:2018oaa}, by  introducing a ultraviolet cutoff $\Lambda$ and by subtracting all the terms that are divergent as $\Lambda\to\infty$. As we have discussed in~\cite{Adshead:2018oaa}, we expect the result of this procedure to be equivalent to that obtained by adiabatic subtraction in the limit $\xi\gg 1$.

After long calculations, which we outline in appendix~\ref{ap:quartic}, we obtain the leading contribution from the quartic diagrams
\begin{align}
\sum_{\mathrm{vertices}} \left< \hat{\gamma} _{{\bf p}_1 }^{\lambda_1}(\tau) \hat{\gamma} _{{\bf p}_2}^{\lambda_2}(\tau) \right>_{\mathrm{quartic}}
&\simeq -\frac{8\, H^4 \log (-p_1\tau)}{9 \pi\,  M_{\rm {Pl}}^4\, p_1^3} \,\mu ^2\, \xi ^3 \,\delta({\bf p}_1 + {\bf p}_2 )\, \delta^{\lambda_1,\lambda_2}\,,
\label{quartic-result}
\end{align}
in the limit $\mu\ll 1\ll \xi$ and for superhorizon modes $-k\tau\ll 1$.
We note that this contribution is parity-even. Parity-odd terms are associated to the operators $H_{\mathrm{int},2}^{(4)}$, $H_{\mathrm{int},3}^{(4)}$ and $H_{\mathrm{int},5}^{(4)}$, which contain the Levi-Civita symbol. However, the contributions from $H_{\mathrm{int},2}^{(4)}$ and from the parity-odd part of $H_{\mathrm{int},5}^{(4)}$ vanish identically after angular integrations, so that the only parity-odd contribution to the tensor power spectrum is given by $H_{\mathrm{int},3}^{(4)}$ and yields
\begin{align}
\left< \hat{\gamma} _{{\bf p}_1 }^{\lambda_1}(\tau) \hat{\gamma} _{{\bf p}_2}^{\lambda_2}(\tau) \right>_{\mathrm{parity-odd}}
&\simeq \lambda_1\,\dfrac{ H^4
}{3\,M_{\rm Pl}^4 \,p_1^3}\, \mu ^2\, \xi ^2\,  \delta({\bf p}_1 + {\bf p}_2 )\, \delta^{\lambda_1, \lambda_2} \,,
\end{align}
which is sub-leading, by a factor $1/\xi$, with respect to the parity-even component.

\subsubsection{Cubic loop}%
\label{sec:cubic_loop}

Next we consider the cubic loop,  shown on the right side of figure \ref{fig:diagrams}. There is a single contribution to this diagram, given by
\begin{align}\label{eq:cubic_inin}
\left\langle\gamma_{\bp_1}^{\lambda_1}(\tau)\,\gamma_{\bp_2}^{\lambda_2}(\tau)\right\rangle^{(3)}=-\frac{2}{M_{\rm Pl}^2\,a(\tau)^2}\int^\tau d\tau_1\int^{\tau_1} d\tau_2\left\langle\left[\left[\t_{\bp_1}^{\lambda_1}(\tau)\,\t_{\bp_2}^{\lambda_2}(\tau),\,H^{(3)}_{\rm int}(\tau_1)\right],\,H^{(3)}_{\rm int}(\tau_2)\right]\right\rangle\,,
\end{align} 
where $H^{(3)}_{\rm int}(\tau)$ is given by eq.~(\ref{H3}). Unlike those appearing in the quartic loops, the integrals in the cubic loop are prohibitively difficult to evaluate exactly.  The expressions appearing in this diagram, however, are very similar to those which appeared in the cubic loop contribution to the spectrum of scalar perturbations considered in reference~\cite{Adshead:2018oaa}.  Therefore, we apply the same sequence of approximations developed in that work to the present calculation.  Here we outline these approximations; the details of the calculation are presented in appendix~\ref{ap:cubic}.

We start by setting the external momenta to zero; as discussed in~\cite{Adshead:2018oaa} we expect this approximation to generate at most a ${\cal O}(1)$ error. Next, since the functions appearing in the integrals are rapidly oscillating, we perform a Wick rotation on the time integration variables, so that the Whittaker functions appearing in the fermion mode functions now have real argument and are exponentially increasing or decreasing.  Next, we approximate those Whittaker functions as linear combinations of monomials times exponentials, with special attention given to the branch cuts. The explicit form of those approximations are given in eqs.~(\ref{eq:Positive_Axis}) and~(\ref{eq:Negative_Axis}), and we have verified their validity in the regime $\xi\gg 1$ we are interested in. These approximate expressions contain a part that behaves like positive frequency (we schematically denote the coefficient of this part by ${\cal A}$) and a part that behaves like negative frequency (whose coefficient is denoted schematically by ${\cal B}$).  Explicit expressions for $\cal{A}$ and $\cal{B}$ can be found in appendix~\ref{ap:cubic}. 

Once the above approximations are in place, the integrals can be computed analytically. We find a divergence in the limit $\tau_2\to\tau_1$, although it is only present in the ${\cal A}^2$ term. Since this term corresponds to positive frequency, ``vacuum only'' modes, we subtract them. Once this component is subtracted, we are left with the final result 
\begin{align}
\left< \hat{\gamma}^{\lambda_1}_{{\bf p}_1}(\tau)
\hat{\gamma}^{\lambda_2}_{{\bf p}_2}(\tau) \right>^{(3)}
& \sim  \mathcal{O}(0.1) \times \dfrac{H^4 \delta({\bf p}_1 + {\bf p}_2)\delta^{\lambda_1,\lambda_2}}{M_{\rm {Pl}}^4 \, p_1^3}  \mu ^2 \,\xi ^3\, \log(-p_1\tau), 
\label{cubic-result}
\end{align}
which has the same parametric dependence as the contribution from the quartic loop.

\subsection{Scaling of our result}
\label{sec:discussion}%

The gravitational wave power spectrum is related to the two point function as 
\begin{align}
\langle\gamma^{\lambda_1}_{\bp_1}(\tau)\,\gamma^{\lambda_2}_{\bp_2}(\tau)\rangle=\frac{2\pi^2}{p_1^3}\, {\cal P}_t^\lambda\,\delta(\bp_1+\bp_2)\,\delta^{\lambda_1,\lambda_2} \;. 
\end{align}

The contribution from the produced fermions, using the results eqs.~(\ref{quartic-result}) and (\ref{cubic-result}) we derived in the previous subsections, is
\begin{align}\label{eq:final_deltapt}
\delta P_t^\lambda\simeq {\cal O}(0.01)\,\frac{H^4}{M_{\rm Pl}^4}\mu^2\,\xi^3\, \log(-p_1\tau).
\end{align}

We now compare this sourced gravitational wave signal to the vacuum contribution, ${\cal P}_t^{\rm {vacuum}}\sim 0.1\,H^2/M_{\rm Pl}^2$.  Their ratio can be written as 
\begin{align}
\frac{\delta{\cal P}_t}{{\cal P}_t^{\rm vacuum}}\simeq 0.1\,\frac{\mu^2\,\xi^3\,H^2}{M_{\rm Pl}^2}\, \log(-p_1\tau)\,.
\end{align}
As discussed in reference~\cite{Adshead:2018oaa}, the quantity $\mu^2\,\xi^3\,H^2/M_{\rm Pl}^2$ corresponds to the ratio between the energy density in fermions and the total energy density in the Universe, which must be much smaller than unity.  We thus conclude that an axion-like inflaton coupled to fermions cannot produce tensor modes that dominate over the spectrum of vacuum fluctuations, in contrast to the scenario in which an axionic inflaton is coupled to gauge fields.

Finally, in reference~\cite{Adshead:2018oaa} we determined the overall scaling, as a function of the parameter $\xi$ and $\mu$, of the diagrams that are relevant for the scalar (bi)spectrum. Here we apply analogous arguments to the diagrams that led to the result in eq.~(\ref{eq:final_deltapt}) for  the tensor spectrum.

Our first observation is that the cubic interaction~\eqref{S3F} gives a contribution $\sim \gamma_{ij}\,\bar\psi\,\Gamma\,p\,\psi$, whereas each quartic interaction in eq.~(\ref{S4F}) gives a contribution $\sim \gamma_{ij}^2\,\bar\psi\,\Gamma\,p\,\psi$, where $p$ schematically denotes a quantity that scales as fermion momentum and $\Gamma$ denotes some combination of the Dirac $\gamma$-matrices (exceptions are ${\cal L}_2^{(4)}$ and ${\cal L}_3^{(4)}$ which contain no dependence on the fermion momentum, but only on the graviton momentum).  The contribution from $\langle\bar\psi\,\gamma^i\,\hat{p}_i\,\gamma^5\,\psi\rangle$ is ultimately zero, due to the asymmetry of the Levi-Civita symbol. A numerical evaluation shows that the  combinations $\langle\bar\psi\,\gamma^i\,\hat{p}_i\,\psi\rangle$ and $\langle\bar\psi\,\gamma^5\,\gamma^0\,\psi\rangle$ (which appear in both the cubic and quartic gravitational vertices) oscillate with amplitude $\mu^2/\xi$, for momenta up to $-k\tau\simeq \xi$. Moreover, the fermion part of the operators in ${\cal L}^{(4)}_{4,\,5,\,6,\,7}$ can always be brought to a form $\langle{\Im}\left(\partial_k\bar\psi\,\gamma^0\psi\right)\rangle$ (see eqs.~(\ref{eq:c3}) and~(\ref{eq:c4}) --  here ${\Im}$ denotes the imaginary part), which can also be seen to scale as $\mu^2/\xi$  for momenta up to $-k\tau\simeq \xi$. (We note for comparison that the bilinears which appear in the diagrams involving fluctuations of the inflaton, $\langle\bar\psi\,\psi\rangle$ and $\langle\bar\psi\,\gamma^5\psi\rangle$, oscillate instead with amplitude $\mu/\xi$. These were considered in reference~\cite{Adshead:2018oaa}.)  Since all fermion bilinears in the $\gamma\bar\psi\psi$ and the $\gamma\gamma\bar\psi\psi$ sector scale as $\mu^2/\xi$, each fermionic line in the diagrams of figure~\ref{fig:diagrams} contributes $\mu^2/\xi$.  

We recall that each interaction Hamiltonian (with the exception of ${\cal L}_2^{(4)}$ and ${\cal L}_3^{(4)}$, as noted above) also carries a power of $p$, and therefore each vertex gives an additional power of $\xi$. Furthermore, every fermionic loop integral, which goes as $d^3k$, gives a contribution $\sim\xi^3$. 

Once we apply these scalings to the quartic diagrams, we have a scaling $(\mu^2/\xi)$ (one fermion line) times $\xi$ (one vertex) times $\xi^3$, giving an overall scaling $(\mu^2/\xi)\times(\xi)\times(\xi^3)\sim \mu^2\,\xi^3$. The quartic diagrams involving ${\cal L}_2^{(4)}$ and ${\cal L}_3^{(4)}$ are respectively vanishing (as a consequence of the symmetries of the operator) and scaling as  $(\mu^2/\xi)\times (1)\times(\xi^3)\sim \mu^2\,\xi^2$---the factor of $(1)$ instead of $(\xi)$ originates from the fact that this vertex does not contain a power of the fermion momentum. These results are in agreement with the direct calculations presented in appendix~\ref{ap:quartic}.

For the cubic gravitational diagram, a naive implementation of these scalings would read $(\mu^2/\xi)^2$ (two fermion lines) times $\xi^2$ (two vertices) times $\xi^3$, giving an overall scaling $\sim \mu^4\,\xi^3$. This would disagree by a factor of $\mu^2$ with the result obtained by the direct (albeit approximate) calculation of appendix~\ref{ap:cubic}. The different scaling with $\mu$  can, however, be understood as follows. Each fermion line in fact scales as $\mathcal{O}(1) + \mathcal{O}(\mu^2)$, although in the quartic diagrams the $\mathcal{O}(1)$ contribution is always divergent and therefore is removed by regularization.  The cubic diagram, with two fermion lines, has terms of order $\mathcal{O}(1)$, $\mathcal{O}(\mu^2)$, and $\mathcal{O}(\mu^4)$. The first is again removed by renormalization, leaving the $\mathcal{O}(\mu^2)$ term which arises from interference.

Given the parity violating nature of the system, one can expect a parity violating tensor spectrum $\delta{\cal P}_t^{+1}\neq \delta{\cal P}_t^{-1}$, which we did indeed find.  As mentioned above, however, this is subdominant by a factor $1/\xi$ with respect to the parity-even part.

\section{Conclusion}
\label{sec:conclusion}%

Axion, or natural inflation is a class of models for slow-roll inflation where the required flatness of the potential is protected from radiative corrections by an approximate shift symmetry. This shift symmetry means that any axion-matter couplings must be via derivatives, and the lowest dimension
couplings of an axion inflaton to gauge fields and fermions are given by eq. (\ref{phi-A-psi}). These couplings are typically employed for reheating in these models. The recent literature (see \cite{Pajer:2013fsa} for a review) has shown that the coupling of the axion to gauge fields can lead to a rich phenomenology during inflation. Analogous studies for fermions are more scarce~\cite{Adshead:2015kza,Anber:2016yqr,Adshead:2018oaa,Domcke:2018eki}. The present work is a direct continuation of  our previous paper  \cite{Adshead:2018oaa}, where it was pointed out that it is more convenient to perform computations in the 
redefined basis ${\rm e}^{i \gamma_5 \frac{\varphi}{f}} \, X$ (where $\varphi$ and $X$ are, respectively, the inflaton and the fermionic fields appearing in eq.\ (\ref{phi-A-psi})), as perturbation theory in the $X$ basis breaks down as the fermion mass $m$ approaches zero.  Our previous work considered the non-Gaussianity of inflaton perturbations sourced by the coupled fields.  In the fermionic case, Pauli blocking limits the occupation number in each individual mode, and therefore phenomenologically interesting results occur only when  the Fermi sphere of produced modes becomes large.  The many modes produced are uncorrelated sources of inflaton perturbations, and therefore summing over the modes produces Gaussian inflaton perturbations by the central limit theorem.  This is in contrast  to the scenario in which the axion inflaton is coupled to gauge fields. In that case, the inflaton perturbations of any given scale $k$ are mostly produced by a few (highly amplified) gauge field modes of comparable scale, thus resulting in a strongly non-Gaussian sourced signal \cite{Barnaby:2010vf}. 

In this paper we complemented our previous study with the computation of the sourced gravitational waves.  We worked in the ADM basis, in which the nondynamical metric perturbations $N$ and $N^i$ are integrated out via the corresponding energy and momentum constraints.  We have solved these constraints perturbatively in the fermionic field $\psi$ and in the gravitational waves $\gamma$. We thus obtained the ${\cal O } \left(  \gamma\bar\Psi\Psi \right)$ and  ${\cal O } \left( \gamma\gamma\bar\Psi\Psi  \right)$ interactions, which we used to compute the contribution to the gravitational wave spectrum from the diagrams shown in figure~\ref{fig:diagrams}. The left diagram is technically simpler; after regularizing it we were able to evaluate it exactly, as described in appendix~\ref{ap:quartic}. The right diagram is much more involved, necessitating the approximations discussed in appendix~\ref{ap:cubic}, analogous to those made for the cubic diagram in reference~\cite{Adshead:2018oaa}.  Although the computations were very involved, in section~\ref{sec:discussion} we presented some simple scaling arguments that correctly capture the scaling of the result with the parameters, $\mu$ and $\xi$, of the model.
 
Our main conclusion is that, in contrast to the scenario in which the axion inflaton is coupled to vector fields, the gravitational waves sourced by the fermions cannot be greater than the vacuum gravitational waves. This conclusion holds also in the regime, studied in reference~\cite{Adshead:2018oaa}, where the strong backreaction of the fermion degrees of freedom controls the dynamic of the zero mode of the inflaton, and we plan to further explore the implications of our studies for inflationary model building.

\acknowledgments

The work of P.A. and L.P. was supported by the US Department of Energy through grant DE-SC0015655.  L.P. gratefully acknowledges support from a Fortner Fellowship at the University of Illinois at Urbana-Champaign. L.P. thanks the Aspen Center for Physics for hospitality and support
through National Science Foundation grant PHY-1607611.  The work of M.R. and L.S. is partially supported by the US-NSF grant PHY-1820675. 

\appendix

\section{Computation of the fermion-gravitational wave interactions}%
\label{ap:lagrangian}

In this appendix we discuss the steps that lead from eq. (\ref{fermion-action1}) to eq. (\ref{fermion-action21}). We begin by writing down the vielbeins for the line element in  eq.\ \eqref{line}, $g_{\mu\nu} = \eta_{AB}e^{A}{}_{\mu}e^{B}{}_{\nu}$.  These are given by
\begin{align}
e^A \phantom{}_\mu
 &= \begin{pmatrix}
N & 0 \\
N^i e^a \phantom{}_i & e^a \phantom{}_i
\end{pmatrix}\,, 
\label{vierbein}
\end{align} 
where the spatial components $e^a \phantom{}_i$ satisfy $\delta_{ab}e^a \phantom{}_ie^b \phantom{}_j=h_{ij}$. Starting from (\ref{vierbein}), we can also write 
\begin{eqnarray} \label{eqn:morevielbein}
e_{A\mu} = \eta_{AB}e^B{}_\mu  
= \(\begin{matrix} N & 0 \\ N^i \eta_{ab}e^{b}{}_{i} & \eta_{ab}e^{b}{}_{j} \end{matrix}\) \;,\;\; 
e_A{}^\mu =  \(\begin{matrix} \frac{1}{N} &  -\frac{N^j}{N} \\ 0 & e_a{}^{j} \end{matrix}\) \;,\;\; 
e^{A\mu} =   \(\begin{matrix} \frac{1}{N} &  -\frac{N^j}{N} \\ 0 & \eta^{ab}e_{b}{}^{j} \end{matrix}\) \;,
\end{eqnarray} 
and one can indeed verify that the product $g^{\mu \nu} = e_A^\mu e_B^\nu \eta^{AB}$ is the inverse of the metric in eq.~(\ref{line}). 

In terms of the vielbein, the spin connections are obtained from 
\begin{align}
\omega_{\mu}{}^{AB} = e^A{}_{\nu}\nabla_\mu e^{B\nu} = e^A{}_{\nu}\(\partial_\mu e^{B\nu} +\Gamma^\nu{}_{\sigma \mu}e^{B\sigma}\) \;,
\end{align}
where $\Gamma^\nu{}_{\sigma \mu}$ are the usual Christoffel symbols associated with the metric $g_{\mu\nu}$. Using eq.\ \eqref{vierbein} and \eqref{eqn:morevielbein}, we find the components of the spin connection in ADM coordinates
\begin{eqnarray} 
\omega_{0}{}^{0b} &=&  \(\partial_i N - N^j K_{ij}\)\eta^{ab}e_a{}^{i} \;, \nonumber\\ 
\omega_{0}{}^{ab} & = &  e^a{}_{k}\partial_0 e_{c}{}^{k}\eta^{bc} +\( - N h^{ik} K_{km} + ^{(3)}\!\nabla_m N^i\)e^a{}_{i}e_{c}{}^{m}\eta^{bc} \;, \nonumber\\ 
\omega_{i}{}^{0b} & = & - K_{ki}e_{c}{}^{k}\eta^{bc} \;, \nonumber\\ 
\omega_{i}{}^{ab} & = &  e^a{}_{k}\partial_i e_{c}{}^{k}\eta^{bc}  - ^{(3)}\!\Gamma^m{}_{k i}e^a{}_{m}e_{c}{}^{k}\eta^{bc} \;. 
\end{eqnarray}
Inserting these in the action (\ref{fermion-action1}) we obtain an expanded form of the fermion action in ADM coordinates
\begin{align}\label{eqn:appferm}
S_{F} = \int d^4 x \, \mathcal{L}_F\,,
\end{align}
where
\begin{align}\nn
\mathcal{L}_F =    a^3 \Bigg\{  &  i\bar\Psi \gamma^{0} \Big[ \partial_{0}+  \(\partial_i N  - N^j \, K_{ij}\)e_b{}^{i}\Sigma^{0b}  +\frac{1}{2}e^{c}{}_{k}\(\partial_0 e_b{}^{k} -\(   N K^{k}{}_{m} - ^{(3)}\!\nabla_m N^k\)e_b{}^{m}\)\eta_{ac}\Sigma^{ab} \Big] \Psi  \\\nn
&  + i\bar\Psi \(\gamma^{a}Ne_{a}{}^k-\gamma^{0}N^k\)\[\partial_{k} - K_{ik} e_b{}^{i}\Sigma^{0b}+\frac{1}{2}\(e^{c}{}_{i}\partial_k e_b{}^{i}  +^{(3)}\!\Gamma^m{}_{ik}e^{c}{}_{m}e_b{}^{i}\)\eta_{ac}\Sigma^{ab} \] \Psi
 \\ & - N m\bar{\Psi}\[ \cos\(\frac{2\varphi}{f}\)+i \sin\(\frac{2\varphi}{f}\)\gamma_5\]\Psi\Bigg\} \, ,
\label{fermion-action2} 
\end{align}
and $\Sigma^{AB} = [\gamma^A, \gamma^B]/4$. The full action is the sum of the bosonic part in eq.\ \eqref{S-bose} and the fermionic part in eq.\ \eqref{eqn:appferm}.

We are interested in the interactions between the fermions and the tensor modes of the metric, and so we expand the spatial part of the vielbein as
\begin{align}
e^a{}_{i} =   a\, \delta^{ak}e^{\frac{1}{2}\gamma_{ki}} &=   a\, \delta^{ak}\[\delta_{ki} + \frac{1}{2}\gamma_{ki}+\frac{1}{8}\gamma_{kj}\gamma_{ji}+\ldots\] \;, \\
e_a{}^{i} =  a^{-1}\, \delta^k_a \delta^{ij} e^{-\frac{1}{2}\gamma_{kj}} &=  a^{-1}\,  \delta^k_a \delta^{ij} ,\[\delta_{ij} - \frac{1}{2}\gamma_{ij}+\frac{1}{8}\gamma_{i\ell}\gamma_{\ell j}+\ldots\] \;, 
\end{align}
which leads to the following components of the spin connection (expanded here to quadratic order in tensors) 
\begin{align}
\omega_{0}{}^{0b} = & a^{-1} \(\partial_a N  + \frac{\mathcal{H}}{N} N_a \)\eta^{ab} \;\;, \\
\omega_{0}{}^{ab}
 =  &- \frac{1}{8}\eta^{ac}\eta^{bd}\(\gamma_{cj}'\gamma_{jd}  - \gamma_{cj}\gamma_{jd}'\) -\frac{a^{-2}}{2}\(\partial_{d}N_{c}-\partial_{c}N_{d}\)\eta^{ac}\eta^{bd} \;\;, \\
\omega_{i}{}^{0b}  = & \frac{a \mathcal{H} }{N}\eta^{bc}\(\delta_{ci}+\frac{1}{2}\gamma_{ci}+\frac{1}{8}\gamma^2_{ci}\) -\frac{1}{2aN}\eta^{bc}\(\partial_{i}N_{c}+\partial_{c}N_{i}\) +\frac{a}{2N}\eta^{bc}\(\gamma_{ic}'+\frac{1}{2} \gamma_{in}\gamma_{nc}'\) \;\;, \\ 
\omega_{i}{}^{ab}= &-\eta^{ac}\eta^{bd}\Big(\frac{1}{2}(\partial_{d}{\gamma_{ic}}  -\partial_{c}{\gamma_{id}})+ \frac{1}{8}\(\gamma_{dk}\partial_i\gamma_{kc} - \gamma_{ck}\partial_i\gamma_{kd}\) \nonumber\\ 
 &+ \frac{1}{4}(\partial_{d}\gamma^2_{ic} -\partial_{c}\gamma^2_{id})+\frac{1}{4}\(\gamma_{dk}\partial_{c}{\gamma_{ki}} - \gamma_{cm}\partial_{d}{\gamma_{mi}}\)  +\frac{1}{4}\( \gamma_{ck}\partial_{k}{\gamma_{id}}-  \gamma_{dk}  \partial_{k}{\gamma_{ic}} \)\Big) \;. 
\end{align}
These relations, along with the expansion of the lapse and shift in eqs.\ \eqref{al-be-th},  are inserted into the constraint equations (\ref{constr1}) and  (\ref{constr2}). The part of the Hamiltonian constraint equation quadratic in field fluctuations is
\begin{align} \label{eqn:HamilQuad}
& 4 \mathcal{H} M_{\rm {Pl}}^2 \Delta \theta^{(2)}
+ \dfrac{M_{\rm {Pl}}^2}{4} \left[ (\gamma^\prime)_{ij} (\gamma^\prime)_{ij}
+  \left( \partial_j \gamma_{kq}  \right)\partial_j \gamma_{qk}  \right] 
 \nonumber \\
& \qquad  =  
- 4 a^2  \alpha^{(2)}  V(\varphi_0)
 +  2a  \left[ \dfrac{i}{2} 
 \left(  \bar{\Psi} \gamma^a \partial_a \Psi
 - (\partial_a  \bar{\Psi}) \gamma^a \Psi \right)
- m \bar{\Psi} \left[
\cos  \left( \dfrac{2\varphi_0}{f} \right) - i \gamma^5 \sin \left( \dfrac{2 \varphi_0}{f} \right)\right] \Psi
\right] ,
\end{align}
while the quadratic part of the momentum constraint reads
\begin{align} \label{eqn:momquad}
0 &= 2
 M_{\rm {Pl}}^2 \mathcal{H} \, \partial_j \alpha^{(2)}
+  M_{\rm {Pl}}^2\left[
- \dfrac{1}{2} \Delta \beta_j^{(2)}
- \dfrac{1}{4} (\partial_i \gamma^\prime)_{jk} \gamma_{ki}
- \dfrac{1}{4} (\partial_j \gamma_{\ell i}) (\gamma^\prime)_i \phantom{}_\ell 
+ \dfrac{1}{4} (\partial_i \gamma_{jk}) (\gamma^\prime)_{ki}
 \right] \nonumber \\
& \qquad
- a \left[\dfrac{i}{2} \left( 
\bar{\Psi} \gamma^0 \partial_j \Psi
- ( \partial_j \bar{\Psi}) \gamma^0  \Psi \right) 
- \dfrac{1}{4}  
   \epsilon_{jab} 
\partial_a  \left(  \bar{\Psi} \gamma^b \gamma^5 \Psi \right) \right]\,.
\end{align} 
Eqs.\ \eqref{eqn:HamilQuad} and \eqref{eqn:momquad} can be solved to find the quadratic order lapse and shift
\begin{align}
 \alpha^{(2)}  
&= \Delta^{-1} \left\lbrace \dfrac{1}{8 \mathcal{H}} \partial_j \left[
(\partial_j \gamma_{\ell i}) (\gamma^\prime)_{i \ell} 
 \right] 
+ 
\dfrac{ia}{4  M_{\rm {Pl}}^2  \mathcal{H}}
 \left[
\bar{\Psi} \gamma^0 \Delta \Psi
- (  \Delta \bar{\Psi}) \gamma^0 \Psi
 \right] \right\rbrace, \nonumber \\
\beta_j^{(2)}
&= \Delta^{-1} \left\lbrace
\dfrac{1}{2} \Delta^{-1}  \partial_j \partial_k \left[
(\partial_k \gamma_{\ell i}) (\gamma^\prime)_{i \ell}
 \right] 
 - \dfrac{1}{2} \left[
(\partial_i \gamma^\prime)_{jk} \gamma_{ki}
+(\partial_j \gamma_{\ell i}) (\gamma^\prime)_{i \ell}
-(\partial_i \gamma_{jk}) (\gamma^\prime)_{ki}
\right]
\right. \nonumber \\
& \qquad \left.
+ 
\dfrac{ia}{M_{\rm {Pl}}^2 }
\partial_j \Delta^{-1} \left[
\bar{\Psi} \gamma^0 \Delta \Psi
- (\Delta \bar{\Psi}) \gamma^0 \Psi
 \right] \right. \nonumber \\
& \qquad \left.
- \dfrac{a}{M_{\rm {Pl}}^2} \left[i \left( \bar{\Psi} \gamma^0 \partial_j \Psi - (\partial_j \bar{\Psi}) \gamma^0 \Psi \right)
- \dfrac{1}{2}  \epsilon_{ijk } \partial_i (\bar{\Psi} \gamma^k \gamma^5 \Psi) 
\right]
\right\rbrace ,\nonumber \\
\theta^{(2)}
&= \Delta^{-1} \left\lbrace - 
 \dfrac{1}{16 \mathcal{H}} \left[ (\gamma^\prime)_{ij} (\gamma^\prime)_{ij}
+  \left( \partial_j \gamma_{kq}  \right)\partial_j \gamma_{qk}  \right] 
-  \dfrac{ia}{4  M_{\rm {Pl}}^2 \mathcal H}   
\left( \bar{\Psi} \gamma^0 \partial_0 \Psi - (\partial_0 \bar{\Psi}) \gamma^0  \Psi \right) \right. \nonumber \\
& \qquad \left.
-  \dfrac{a^2}{M_{\rm {Pl}}^2 \mathcal H} V( \varphi_0) 
\Delta^{-1} \left\lbrace \dfrac{1}{8 \mathcal{H}} \partial_j \left[
(\partial_j \gamma_{\ell i}) (\gamma^\prime)_{i \ell}
 \right] 
+ 
\dfrac{ia }{4  M_{\rm {Pl}}^2  \mathcal{H}}
 \left[
\bar{\Psi} \gamma^0 \Delta \Psi
- (  \Delta \bar{\Psi}) \gamma^0 \Psi
 \right] \right\rbrace   \right\rbrace,
\label{constsol-2_app}
\end{align}
where $\Delta = \partial_i\partial_i$ is the spatial Laplacian, and $\Delta^{-1}$ is its inverse. In deriving these solutions, we have disregarded fluctuations of the inflaton field because we are only interested in the interactions between  gravitational waves and  fermions. The inclusion of inflaton fluctuations introduces terms that are quadratic in the inflaton fluctuations, as well as terms quadratic in the first order perturbation to the lapse and shift.\footnote{Our solutions for the part of the second order constraints that is quadratic in tensors differs from the results presented in  reference \cite{Dimastrogiovanni:2008af}; however, this does not affect the conclusions of that work.} We have also made use of the linear order equation of motion for the fermion.  This induces corrections to the action that begin at fifth order in fluctuations and are thus irrelevant here.

We are now ready to evaluate the action, eq.\ (\ref{fermion-action2}), on the constraint surface and eliminate the non-dynamical lapse and shift. Inserting the solutions to the constraints \eqref{constsol-2_app} into the full action, eq.\ \eqref{S-bose} + \eqref{fermion-action21}, we expand the result to quartic order. This results in an action for the dynamical fields, $\psi$ and $\gamma$, consisting of a quadratic (free) part, $S^{(2)}$ and cubic and quartic parts, $S_F^{(3)}$ and $S_F^{(4)}$, which describe the interactions of a fermion bilinear with one and two gravitational waves, respectively. The action that results from the procedure just described is very long and complicated. However, in practice there are a number of facts that considerably simplify the result. Because the constraint equations are derived from the variation of the action with respect to the lapse and shift, one can use their equations of motion before substituting in their solutions. This results in the cancellation of a large number of terms, and leaves the result we report above in eqs.\ \eqref{S3F} and \eqref{S4F}.

\section{Interaction Hamiltonian}
\label{ap:Hint}

In this appendix we write the explicit forms of the interaction Hamiltonian terms eq.\ (\ref{Hint}), obtained  by inserting the decompositions~(\ref{gamma-deco1}) and~(\ref{psi-deco1}) into the actions in eqs.\ (\ref{S3F}) and (\ref{S4F}). For the cubic term, we find 
\begin{align}
H^{(3)}_{\rm int}  =  - \frac{1}{2\sqrt{2}M_{\rm Pl}} \frac{1}{a(\tau)}   \sum_{\lambda} \int \frac{\prod_{i = 1}^3d^3 k_i}{(2\pi)^{3/2}} \t^{\lambda}_{{\bf k}_1} \bar\psi_{{\bf k}_2} \gamma^{c}\psi_{{\bf k}_3}\Pi^{\lambda}_{cj}({\bf k}_1)({\bf k}_2+{\bf k}_3)_j\, \delta^{(3)}({\bf k_1}-{\bf k}_2+{\bf k}_3) \;. 
\label{H3}
\end{align}

The various terms contributing at quartic order are 
\begin{align}
\label{eq:quartic_vertices}
H^{(4)}_{{\rm int},1}  = &   \frac{1}{8 M_{\rm Pl}^2} \frac{1}{a^2} \sum_{\lambda\lambda'} \int \frac{\prod_{i = 1}^4d^3 k_i}{(2\pi)^{3}}
  \t_{{\bf k}_1}^\lambda  \t_{{\bf k}_2}^{\lambda'} \bar\psi_{{\bf k}_3} \gamma^{c}\psi_{{\bf k}_4} \Pi^{\lambda}_{cm}({\bf k}_1)\Pi^{\lambda'}_{mj}({\bf k}_2)({\bf k}_3+{\bf k}_4)_j \,\; \nonumber\\ & \qquad \times \delta^{(3)}({\bf k_1}+{\bf k}_2-{\bf k}_3+{\bf k}_4) \;, \nonumber\\ 
H^{(4)}_{{\rm int},2} = &\frac{1}{8M_{\rm Pl}^2} \frac{1}{a} \sum_{\lambda\lambda'} \int \frac{\prod_{i = 1}^4d^3 k_i}{(2\pi)^{3}}
 \Bigg[\( \frac{\t_{{\bf k}_1}^\lambda}{a}\)' \t_{{\bf k}_2}^{\lambda'} \bar\psi_{{\bf k}_3} \gamma^c \gamma^5\psi_{{\bf k}_4}  \epsilon^{abc} \Pi^{\lambda}_{aj}({\bf k}_1)\Pi^{\lambda'}_{jb}({\bf k}_2) \Bigg]\nonumber\\ & \qquad \times\delta^{(3)}({\bf k_1}+{\bf k}_2-{\bf k}_3+{\bf k}_4)\;, \nonumber\\ 
H^{(4)}_{{\rm int},3} = & \frac{1}{8 M_{\rm Pl}^2} \frac{1}{a^2} \sum_{\lambda\lambda'} \int \frac{\prod_{i = 1}^4d^3 k_i}{(2\pi)^{3}} 
\Bigg[\t^\lambda_{{\bf k}_1}\t_{{\bf k}_2}^{\lambda'} \bar{\psi}_{{\bf k}_3}\gamma^0 \gamma^5\psi_{{\bf k}_4}   \lambda' {k}_{2} \Pi^{\lambda}_{bk}({\bf k}_1)\Pi^{\lambda'}_{kb}({\bf k}_2) \Bigg]\nonumber\\ & \qquad \times\delta^{(3)}({\bf k_1}+{\bf k}_2-{\bf k}_3+{\bf k}_4)\;, \nonumber\\ 
H^{(4)}_{{\rm int},4} = &- \frac{i}{2\,aM_{\rm Pl}^2} \(1-\frac{V }{4M_{\rm Pl}^2 \cH^2}\) \sum_{\lambda\lambda'} \int \frac{\prod_{i = 1}^4d^3 k_i}{(2\pi)^{3}} 
\Bigg[\t^\lambda_{{\bf k}_1}\,\(\frac{\t^{\lambda'}_{{\bf k}_2}}{a}\)'  \bar\psi_{{\bf k}_3} \gamma^{0}\psi_{{\bf k}_4} (k_3^2 - k_{4}^2)\nonumber\\
&\times\frac{( {\bf k}_{1}+{\bf k}_{2})\cdot{\bf k}_{1}}{|{\bf k}_1+{\bf k}_2|^4}\Pi^{\lambda}_{jk}({\bf k}_{1})\Pi^{\lambda'}_{jk}({\bf k}_{2})\Bigg] \,\delta^{(3)}({\bf k_1}+{\bf k}_2-{\bf k}_3+{\bf k}_4) \;, \nn
\end{align}

\begin{align}
H^{(4)}_{{\rm int},5} = &-\frac{1}{4 M_{\rm Pl}^2} \frac{1}{a}  \sum_{\lambda\lambda'}\int \frac{\prod_{i = 1}^4d^3 k_i}{(2\pi)^{3}} 
\Bigg\{\t_{{\bf k}_1}^{\lambda}\(\frac{\t_{{\bf k}_2}^{\lambda'}}{a}\)'\,\nn\\ & \times \[(\bar\psi_{{\bf k}_3}  \gamma^c \gamma^5\psi_{{\bf k}_4}) \frac{\epsilon^{aic}({\bf k}_3-{\bf k}_4)_a}{|{\bf k}_1+{\bf k}_2|^2} - 2i\frac{({\bf k}_3+{\bf k}_4)_i}{|{\bf k}_1+{\bf k}_2|^2} (\bar\psi_{{\bf k}_3}  \gamma^0 \psi_{{\bf k}_4})  \]\nonumber\\
& \times   \[ ({\bf k}_{1})_j\Pi^{\lambda}_{ik}({\bf k}_{1})\Pi^{\lambda'}_{jk}({\bf k}_{2})-({\bf k}_{2})_{j}\Pi^{\lambda}_{jk}({\bf k}_{1})\Pi^{\lambda'}_{ik}({\bf k}_{2})- ({\bf k}_{1})_{i} \Pi^{\lambda}_{jk}({\bf k}_{1})\Pi^{\lambda'}_{jk}({\bf k}_{2})\] \Bigg\} \nonumber \\ & \times \delta^{(3)} ({\bf k_1}+{\bf k}_2-{\bf k}_3+{\bf k}_4) \;, \nonumber\\ 
H^{(4)}_{{\rm int},6} = & -\frac{ i }{16 M_{\rm Pl}^2\,aH} \sum_{\lambda\lambda'} \int \frac{\prod_{i = 1}^4d^3 k_i}{(2\pi)^{3}} 
\Bigg[  \left(\(\frac{\t^{\lambda}_{{\bf k}_1}}{a}\)'\,\(\frac{\t^{\lambda'}_{{\bf k}_2}}{a}\)'-\frac{{\bf k}_1 \cdot{\bf k}_2}{a^2}\t^{\lambda}_{{\bf k}_1}\t^{\lambda'}_{{\bf k}_2}\right) \nonumber\\ & \qquad \times \bar\psi_{{\bf k}_3} \gamma^{0}\psi_{{\bf k}_4} 
 \frac{ k_3^2-k_4^2}{|{\bf k}_3-{\bf k}_4|^2}\Pi^{\lambda}_{ij}({\bf k}_1)\Pi^{\lambda'}_{ij}({\bf k}_2)\Bigg] 
 \delta^{(3)}({\bf k_1}+{\bf k}_2-{\bf k}_3+{\bf k}_4) \;,  \nonumber\\ 
H^{(4)}_{{\rm int},7} = & \frac{i}{8M_{\rm Pl}^2a^2H} \sum_{\lambda\lambda'} \int \frac{\prod_{i = 1}^4d^3 k_i}{(2\pi)^{3}}  
\Bigg[ \(\bar\psi_{{\bf k}_3} \gamma^{0} \partial_{0}\psi_{{\bf k}_4} -\partial_{0} \bar\psi_{{\bf k}_3} \gamma^{0} \psi_{{\bf k}_4}\)\t^{\lambda}_{{\bf k}_1}\(\frac{\t^{\lambda'}_{{\bf k}_2}}{a}\)' \nonumber\\ & \qquad \times \frac{{\bf k}_1\cdot({{\bf k}_1+{\bf k}_2})}{|{\bf k}_1+{\bf k}_2|^2}\Pi^{\lambda}_{ij}({\bf k}_1)\Pi^{\lambda'}_{ij}({\bf k}_2)\Bigg] \delta^{(3)}({\bf k_1}+{\bf k}_2-{\bf k}_3+{\bf k}_4) \;. 
\end{align}

\section{Details of the quartic loop computation}
\label{ap:quartic}

Having identified the seven quartic vertices in eq.~\eqref{eq:quartic_vertices}, we now proceed to evaluate the left loop diagram of figure \ref{fig:diagrams}, using the mode functions presented in section \ref{sec:mode_functions}. Evaluating the fermion expectation values produces terms of the form $\delta({\bf k}_3 - {\bf k}_4)$, which sets the denominator in several of the vertices to zero.  To ensure mathematically sensible results, we follow an approach similar to that of~\cite{Seery:2007we}, in which we set ${\bf k}_4 = {\bf k}_3 + \boldsymbol{\epsilon}$ (and therefore ${\bf p}_2 = - {\bf p}_1 +\boldsymbol{\epsilon}$) and take the limit $\boldsymbol{\epsilon} \rightarrow 0$, as discussed below.

It is convenient to note that, to quadratic order in $\boldsymbol{\epsilon}$,  
\begin{align}
 \Pi^{\lambda_1}_{ij}({\bf p})  \Pi^{i j,\, \lambda_2}(-{\bf p} +\boldsymbol{\epsilon}) 
&\simeq 2 \delta^{\lambda_1,\lambda_2}
 \left[ 1 + \dfrac{({\bf p} \cdot \boldsymbol{\epsilon})^2 - p^2 \, {\epsilon}^2}{2 \,p^4} \right] ,
\label{eq:spinor_drop}
\end{align}
which can be found using the explicit expressions for the spinors given in~\cite{Shiraishi:2010kd}; the lack of a linear term allows for several derivations to be simplified.  Similarly, using explicit expressions for spinors one can show that 
\begin{align}
\chi_{-r}^{\dagger}({\bf p}) \chi_{-r}({\bf p}+\boldsymbol{\epsilon}) &\simeq  1 + \dfrac{({\bf p} \cdot \boldsymbol{\epsilon})^2 -  p^2\,{\epsilon}^2}{8 {p}^4}.
\end{align}

We calculate the two point correlation function with a single loop correction; the subscript indicates which of the seven vertices of eq.~(\ref{eq:quartic_vertices}) was used.  Four of the results are explicitly independent of ${\boldsymbol{\epsilon}}$ at lowest order,
\begin{align}\label{eq:c3}
\!\!\!\! \!\!\!\! \!\!\!\! \!\!\!\!
\!\!\!\! \!\!\!\! \!\!\!\! \!\!\!\!
\left\langle \hat \gamma _{{\bf p}_1}^{\lambda_1}(\tau) \hat \gamma _{{\bf p}_2}^{\lambda_2}(\tau) \right\rangle_{1}
&= i\,\dfrac{\delta({\bf p}_1 + {\bf p}_2) \delta^{\lambda_1,\lambda_2}}{2 \,M_{\rm Pl}^4\, a(\tau)^2} 
\int_{-\infty}^\tau \dfrac{d\tau_1}{a(\tau_1)^2} \, 
 \left( \t^{\lambda_1}_{{p}_1}(\tau)^2  t^{\lambda_1}_{{p}_1}(\tau_1)^{*2 } - {\rm {h.c.}}
 \right)\nonumber \\
& \times
\sum_r \int \dfrac{d^3k_3 }{(2\pi)^3}
\left( k_3 - \dfrac{({\bf k}_3 \cdot {\bf p}_1)^2}{k_3\, p_1^2} \right)
\left[  v^r_{k_3}(\tau_1)^*\,   u^r_{k_3}(\tau_1) +  u^r_{k_3}(\tau_1)^*   v^r_{k_3}(\tau_1)\right], \nonumber  \\
\!\!\!\! \!\!\!\! \!\!\!\! \!\!\!\!
\left< \hat \gamma _{{\bf p}_1 }^{\lambda_1}(\tau) \hat \gamma _{{\bf p}_2}^{\lambda_2}(\tau) \right>_{2}
&= 0. \nonumber \\
\!\!\!\! \!\!\!\! \!\!\!\! \!\!\!\!
\left< \hat \gamma _{{\bf p}_1 }^{\lambda_1}(\tau) \hat \gamma _{{\bf p}_2}^{\lambda_2}(\tau) \right>_{3}
&= i \,\lambda_1\,\dfrac{\delta({\bf p}_1 + {\bf p}_2 ) \delta^{\lambda_1, \lambda_2} 
}{2\, M_{\rm Pl}^4 \,a(\tau)^2} \, p_1
\int_{-\infty}^\tau \dfrac{d\tau_1}{a(\tau_1)^2} \, 
\left( \t^{\lambda_1}_{{p}_1}(\tau)^2  t^{\lambda_1}_{{p}_1}(\tau_1)^{*2 } - {\rm {h.c.}}
 \right)\nonumber \\
& \times 
\sum_r r \int \dfrac{ d^3k_3 }{(2\pi)^3}
\left[  v^r_{k_3}(\tau_1)^*\,   u^r_{k_3}(\tau_1) +  u^r_{k_3}(\tau_1)^*   v^r_{k_3}(\tau_1)\right], \nonumber \\
\left< \hat{\gamma} _{{\bf p}_1 }^{\lambda_1}(\tau) \hat{\gamma} _{{\bf p}_2}^{\lambda_2}(\tau) \right>_{6}
& =  \dfrac{\delta({\bf p}_1 + {\bf p}_2 ) \delta^{\lambda_1, \lambda_2}}{12\, M_{\rm Pl}^4\, a(\tau)^2\,H} \int_{-\infty}^\tau \dfrac{d\tau_1}{a(\tau_1)} \nonumber\\
&\times \left[ \t_{{p}_1}^{\lambda_1} (\tau)^2
\left[\left( \dfrac{\t^{\lambda_1}_{{p}_1}(\tau_1)^*}{a(\tau_1)} \right)^\prime\right]^2
+ \dfrac{p_1^2}{a(\tau_1)^2} 
\t_{{p}_1}^{\lambda_1} (\tau)^2 
\t^{\lambda_1}_{{p}_1}(\tau_1)^{*2}
-{\rm { h.c.}}  \right] \nonumber \\
&\times \sum_r   \int \dfrac{ d^3k_3 }{(2\pi)^3} k_3
\left( 
 {v}^r_{k_3}(\tau_1)\, \partial_{k_3} {v}^r_{k_3}(\tau_1)^{* } 
+  {u}^r_{k_3}(\tau_1)\, \partial_{k_3} {u}^r_{k_3}(\tau_1)^{* } 
- {\rm {h.c.}}\right),
\nonumber \\
\end{align}

However, the other three are not explicitly independent of the direction of ${\boldsymbol{\epsilon}}$ at lowest order,
%
\begin{align}\label{eq:c4}
\!\!\!\!\!\!\!\!\!\!\!\!\!\!\!\!\!\!\!\!\!\!\!\!\!\!\!\!\!\!\!\!
\left< \hat \gamma _{{\bf p}_1 }^{\lambda_1}(\tau) \hat \gamma _{{\bf p}_2}^{\lambda_2}(\tau) \right>_{4}
&= -\frac{1}{3}\dfrac{\delta({\bf p}_1 + {\bf p}_2) \,\delta^{\lambda_1,\lambda_2}}{ M_{\rm Pl}^4\, a(\tau)^2}
\int_{-\infty}^\tau \dfrac{d\tau_1}{a(\tau_1)}  \left( 1 - 
\dfrac{ V}{4\,H^2\, M_{\rm Pl}^2} \right)  \nonumber \\
&  \times 
\sum_r   \int \dfrac{ d^3k_3 }{(2\pi)^3} k_3
\left( 
 {v}^r_{k_3}(\tau_1)\, \partial_{k_3} {v}^r_{k_3}(\tau_1)^{* } 
+  {u}^r_{k_3}(\tau_1)\, \partial_{k_3} {u}^r_{k_3}(\tau_1)^{* } 
- {\rm {h.c.}}\right)\nonumber\\
&\times F_{p_1}^{\lambda_1}(\tau,\,\tau_1;\,\boldsymbol{\epsilon}),  \nonumber \\ 
%
%
%
%
%
%
%
 \!\!\!\! \!\!\!\! \!\!\!\! \!\!\!\!
\left< \hat{\gamma} _{{\bf p}_1}^{\lambda_1}(\tau) \hat{\gamma} _{{\bf p}_2}^{\lambda_2}(\tau) \right>_{5}
& = 
\dfrac{\delta({\bf p}_1 + {\bf p}_2 )\, \delta^{\lambda_1,\lambda_2}}{3 M_{\rm Pl}^4 a(\tau)^2} 
\int_{-\infty}^\tau \dfrac{d\tau_1 }{a(\tau_1)}\, \nonumber \\
&  \times
\sum_r   \int \dfrac{ d^3k_3 }{(2\pi)^3} k_3
\left( 
 {v}^r_{k_3}(\tau_1)\, \partial_{k_3} {v}^r_{k_3}(\tau_1)^{* } 
+  {u}^r_{k_3}(\tau_1)\, \partial_{k_3} {u}^r_{k_3}(\tau_1)^{* } 
- {\rm {h.c.}}\right)\nonumber\\
&\times F_{p_1}^{\lambda_1}(\tau,\,\tau_1;\,\boldsymbol{\epsilon}), 
\nonumber \\
\left< \hat \gamma _{{\bf p}_1 }^{\lambda_1}(\tau) \hat \gamma _{{\bf p}_2}^{\lambda_2}(\tau) \right>_{7}
& =  \dfrac{\delta({\bf p}_1 + {\bf p}_2 )\,\delta^{\lambda_1,\lambda_2}}{4 M_{\rm Pl}^4 a(\tau)^2} 
\int_{-\infty}^\tau \dfrac{d\tau_1}{a(\tau_1)^2\,H} \nonumber \\
&  \times  \sum_r  \int \dfrac{d^3k_3 }{(2\pi)^3}
\left( {u}^r_{k_3}(\tau_1)  u^r_{k_3}(\tau_1)^{*\prime} + {v}^r_{k_3}(\tau_1)  v^r_{k_3}(\tau_1)^{*\prime }
-{\rm {h.c.}}\right)
\nonumber \\ 
&\times F_{p_1}^{\lambda_1}(\tau,\,\tau_1;\,\boldsymbol{\epsilon})\,,
\end{align}
where
\begin{align}
F_{p_1}^{\lambda_1}(\tau,\,\tau_1;\,\boldsymbol{\epsilon})\equiv &\left\lbrace
\t_{p_1}^{\lambda_1}(\tau)^2
\t_{p_1}^{\lambda_1}(\tau_1)^{*}
\left( \dfrac{\t_{p_1}^{\lambda_1}(\tau_1)^*}{a(\tau_1)} \right)^\prime\right.\nonumber\\
&\quad -\left.
\lim_{\boldsymbol{\epsilon} \rightarrow 0} \dfrac{({\boldsymbol{\epsilon}} \cdot {\bf p}_1)^2}{{\epsilon}^2 \,p_1} 
 \,\left[
\t_{p_1}^{\lambda_1}({\tau})
\t_{p_1}^{\lambda_1}(\tau_1)^{*} 
\partial_{p_1} \left[
\t_{p_1}^{\lambda_1}(\tau) 
\left( \dfrac{\t_{p_1}^{\lambda_1}(\tau_1)^{*}}{a(\tau_1)} \right)^\prime  \right]\right.\right.\nonumber\\
&\qquad\left.\left. 
- 
\partial_{p_1} \left[
\t_{p_1}^{\lambda_1}({\tau})\,\t_{p_1}^{\lambda_1}({\tau_1})^*\right]
\,\t_{p_1}^{\lambda_1}({\tau})
\left( \dfrac{\t_{p_1}^{\lambda_1}({\tau}_1)^*}{a(\tau_1)} \right)^\prime  \right]  
- {\rm {h.c.}}    
\right\rbrace\,.
\end{align}
Following reference~\cite{Seery:2007we}, we impose that the limit is approached from an orthogonal direction, ${\bf p}_1 \cdot { \boldsymbol{\epsilon}} \sim {\epsilon}^2$, which makes the problematic terms subdominant.

To separate the integrals, we introduce the new variables $x_1 = - p_1 \tau_1$ and $y_1 = - k_3 \tau_1$; we also take $p_1 \tau \rightarrow 0$, to compute superhorizon quantities.  We arrive at
\begin{align}
\left< \hat \gamma _{{\bf p}_1 }^{\lambda_1}(\tau) \hat \gamma _{{\bf p}_2}^{\lambda_2}(\tau) \right>_{1}
&=- \dfrac{ H^4 \delta({\bf p}_1 + {\bf p}_2) \delta^{\lambda_1,\lambda_2}}{12 \pi^2 M_{\rm Pl}^4 p_1^3} 
\int^{\infty}_x \dfrac{dx_1}{x_1^4} \, 
  \left(\left(x_1^2-1\right) \sin (2 x_1)+ 2 x_1 \cos (2 x_1)\right)  \nonumber \\
 & \times 
 \sum_r  \int dy_1 \, y_1^3 
\left[  v^r(y_1)^*\,   u^r(y_1) +  u^r(y_1)^*   v^r(y_1)\right]\,, \nonumber \\
\left< \hat \gamma _{{\bf p}_1 }^{\lambda_1}(\tau) \hat \gamma _{{\bf p}_2}^{\lambda_2}(\tau) \right>_{2}
&= 0\,,\nonumber \\
\left< \hat \gamma _{{\bf p}_1 }^{\lambda_1}(\tau) \hat \gamma _{{\bf p}_2}^{\lambda_2}(\tau) \right>_{3}
&= -\dfrac{ H^4 \lambda_1 \delta({\bf p}_1 + {\bf p}_2 ) \delta^{\lambda_1, \lambda_2} 
}{8 \pi^2 M_{\rm Pl}^4 p_1^3} 
 \int^{\infty}_x \dfrac{dx_1}{x_1^3} \, 
 \left(\left(x_1^2-1\right) \sin (2 x_1)+2 x_1 \cos (2 x_1)\right)
\nonumber \\
& \times \sum_r r \int dy_1 \, y_1^2
\left[  v^r(y_1)^*\,   u^r(y_1) +  u^r(y_1)^*   v^r(y_1)\right] \,,\nonumber 
\\
\!\!\!\!\!\!\!\!\!\!\!\!\!\!\!\!\!\!\!\!\!\!\!\!\!\!\!\!\!\!\!\!
\left< \hat \gamma _{{\bf p}_1 }^{\lambda_1}(\tau) \hat \gamma _{{\bf p}_2}^{\lambda_2}(\tau) \right>_{4}
&= -\dfrac{i H^4 \delta^{\lambda_1,\lambda_2} \delta({\bf p}_1 + {\bf p}_2)}{12 \pi^2 M_{\rm Pl}^4 p_1^3}
 \left( 1 - 
\dfrac{ V(\varphi_0) }{4  M_{\rm Pl}^2 H^2 } \right)
\int^{\infty}_x \dfrac{dx_1}{x_1^2}  
 [\sin (2 x_1)-x_1 \cos (2 x_1)] 
\nonumber \\ 
&  \times  \sum_r  \int dy_1 \, y_1^3
\left( 
 {v}^r(y_1)\, \partial_{y_1} {v}^r(y_1)^{* } 
+  {u}^r(y_1)\, \partial_{y_1} {u}^r_{k_3}(y_1)^{* } 
- {\rm {h.c.}}\right)\,, \nonumber 
\\
 \!\!\!\! \!\!\!\! \!\!\!\! \!\!\!\!
\left< \hat{\gamma} _{{\bf p}_1 }^{\lambda_1}(\tau) \hat{\gamma} _{{\bf p}_2}^{\lambda_2}(\tau) \right>_{5}
& = 
\dfrac{i H^4 \delta^{\lambda_1,\lambda_2} \delta({\bf p}_1 + {\bf p}_2 )}{12 \pi^2 M_{\rm Pl}^4 p_1^3}  
\int^{\infty}_x \dfrac{dx_1}{x_1^2} \, 
 [\sin (2 x_1)-x_1 \cos (2 x_1)] 
\nonumber \\
&  \times  \sum_r  \int dy_1 \, y_1^3
\left( 
 {v}^r(y_1)\, \partial_{y_1} {v}^r(y_1)^{* } 
+  {u}^r(y_1)\, \partial_{y_1} {u}^r_{k_3}(y_1)^{* } 
- {\rm {h.c.}}\right) \,,\nonumber \\
\!\!\!\! \!\!\!\! \!\!\!\! \!\!\!\!
 \!\!\!\!\!\!\!\!\!\!\!\!\!\!\!\!\!\!\!\!\!\!\!\!\!\!\!\!\!\!\!\!
\left< \hat{\gamma} _{{\bf p}_1 }^{\lambda_1}(\tau) \hat{\gamma} _{{\bf p}_2}^{\lambda_2}(\tau) \right>_{6}
& = -\dfrac{i H^4 \delta({\bf p}_1 + {\bf p}_2 ) \delta^{\lambda_1, \lambda_2}}{48 \pi^2 M_{\rm Pl}^4 p_1^3 } \int^{\infty}_x \dfrac{dx_1}{ x_1^2}   [\sin (2 x_1)-2 x_1 \cos (2 x_1)]  \nonumber \\
&  \times  \sum_r  \int dy_1 \, y_1^3
\left( 
 {v}^r(y_1)\, \partial_{y_1} {v}^r(y_1)^{* } 
+  {u}^r(y_1)\, \partial_{y_1} {u}^r_{k_3}(y_1)^{* } 
- {\rm {h.c.}}\right) \,,\nonumber 
\\
\left< \hat \gamma _{{\bf p}_1 }^{\lambda_1}(\tau) \hat \gamma _{{\bf p}_2}^{\lambda_2}(\tau) \right>_{7}
& = - \dfrac{H^4 \delta^{\lambda_1,\lambda_2} \delta({\bf p}_1 + {\bf p}_2 )}{8 \pi^2 M_{\rm Pl}^4 p_1^3 } 
\int^{\infty}_x \dfrac{dx_1}{x_1^2}  \, 
 [\sin (2 x_1)-x_1 \cos (2 x_1)] \,\sum_r  \int dy_1 \, y_1^2 
 \nonumber \\
&
\!\!\!\!\!\!\!\!\!\!\!\!\times
\left( 
 \dfrac{m_c}{H} (|{u}^r(y_1)|^2 - |{v}^r(y_1)|^2)
+2\, y_1\Re \{{u}^r(y_1){v}^r(y_1)^*\}
-2 r \dfrac{ m_s}{H} \Im \{{u}^r(y_1) {v}^r(y_1)^*\}
\right),
\end{align}
where  $m_c \equiv m \cos(\varphi_0 \slash f)$ and $m_s \equiv m \sin(\varphi_0 \slash f)$.
Several of the $x_1$ integrals are logarithmically divergent and are regulated by the finite amount of e-foldings between when the mode leaves the horizon and the end of inflation.  This logarithmic divergence is a consequence of the fact that the fermions have a nonzero average density  that continues to source graviton fluctuations even when outside the horizon; analogous behavior was observed in the sourced inflaton perturbations of reference~\cite{Adshead:2018oaa}.

The remaining integrals are UV-divergent and need to be regularized. We thus introduce a ultraviolet cutoff $\Lambda$.  (We  discuss renormalization further below.) After some algebra, the expectation values can be expressed in terms of the three integrals
\begin{align}
\mathcal{I}_1 &= \sum_r \int dy_1 \, y_1\, \Re\{s^r(y_1)\,d^r(y_1)^*\}, \nonumber \\
\mathcal{I}_2 &= \sum_r \int dy_1 \, y_1^2\, |s^r(y_1)|^2, \nonumber \\
\mathcal{I}_3 &= \sum_r r \int dy_1 \, y_1\, |s^r(y_1)|^2,
\end{align}
where the functions $s^r(y)$ and $d^r(y)$ are defined in eq.~(\ref{eq:def_uv}). We have
\begin{align}
\left< \hat \gamma _{{\bf p}_1 }^{\lambda_1}(\tau) \hat \gamma _{{\bf p}_2}^{\lambda_2}(\tau) \right>_{1}
&=- \dfrac{ H^4 \delta({\bf p}_1 + {\bf p}_2) \delta^{\lambda_1,\lambda_2}}{9 \pi^2 M_{\rm Pl}^4 p_1^3} 
\cdot  \log(x) 
\left( \mathcal{I}_2 -2  \dfrac{\Lambda^4}{4}
\right),
\nonumber \\
\left< \hat \gamma _{{\bf p}_1 }^{\lambda_1}(\tau) \hat \gamma _{{\bf p}_2}^{\lambda_2}(\tau) \right>_{2}
&= 0, \nonumber \\
\left< \hat \gamma _{{\bf p}_1 }^{\lambda_1}(\tau) \hat \gamma _{{\bf p}_2}^{\lambda_2}(\tau) \right>_{3}
&= \dfrac{ H^4 \lambda_1 \delta({\bf p}_1 + {\bf p}_2 ) \delta^{\lambda_1, \lambda_2} 
}{8 \pi M_{\rm Pl}^4 p_1^3} \mathcal{I}_3,
\nonumber \\
\!\!\!\!\!\!\!\!\!\!\!\!\!\!\!\!\!\!\!\!\!\!\!\!\!\!\!\!\!\!\!\!
\left< \hat \gamma _{{\bf p}_1 }^{\lambda_1}(\tau) \hat \gamma _{{\bf p}_2}^{\lambda_2}(\tau) \right>_{4}
&= -\dfrac{H^4 \delta^{\lambda_1,\lambda_2} \delta({\bf p}_1 + {\bf p}_2)}{3 \pi^2 M_{\rm Pl}^4 p_1^3}
 \left( 1 - 
\dfrac{ V(\varphi_0) }{4  M_{\rm Pl}^2 H^2 } \right)
\log(x) 
\left( - 2 \xi \mathcal{I}_3
- \mu \mathcal{I}_1
+ 2 \dfrac{\Lambda^3}{3} - \mathcal{I}_2 \right),
\nonumber \\
%
 \!\!\!\! \!\!\!\! \!\!\!\! \!\!\!\!
\left< \hat{\gamma} _{{\bf p}_1 }^{\lambda_1}(\tau) \hat{\gamma} _{{\bf p}_2}^{\lambda_2}(\tau) \right>_{5}
& = 
\dfrac{ H^4 \delta^{\lambda_1,\lambda_2} \delta({\bf p}_1 + {\bf p}_2 )}{3 \pi^2 M_{\rm Pl}^4 p_1^3} \log(x) 
\left( - 2 \xi \mathcal{I}_3
- \mu \mathcal{I}_1
+ 2 \dfrac{\Lambda^3}{3} - \mathcal{I}_2 \right), \nonumber \\
\!\!\!\!\!\!\!\!\!\!\!\!\!\!\!\!\!\!\!\!\!\!\!\!\!\!\!\!\!\!\!\!
\left< \hat{\gamma} _{{\bf p}_1 }^{\lambda_1}(\tau) \hat{\gamma} _{{\bf p}_2}^{\lambda_2}(\tau) \right>_{6}
& =  \dfrac{ H^4 \delta({\bf p}_1 + {\bf p}_2 ) \delta^{\lambda_1, \lambda_2}}{6 \pi^2 M_{\rm Pl}^4 p_1^3 }  
\left( - 2 \xi \mathcal{I}_3
- \mu \mathcal{I}_1
+ 2 \dfrac{\Lambda^3}{3} - \mathcal{I}_2 \right),
\nonumber \\
\!\!\!\! \!\!\!\! \!\!\!\! \!\!\!\!
\left< \hat \gamma _{{\bf p}_1 }^{\lambda_1}(\tau) \hat \gamma _{{\bf p}_2}^{\lambda_2}(\tau) \right>_{7}
& =  \dfrac{H^4 \delta^{\lambda_1,\lambda_2} \delta({\bf p}_1 + {\bf p}_2 )}{4 \pi^2 M_{\rm Pl}^4 p_1^3 } \log(x) 
\left( \mu \mathcal{I}_1 
+ \mathcal{I}_2 - 2 \dfrac{\Lambda^4}{4} 
\right).
\label{eq:before_whittaker}
\end{align}

In reference~\cite{Adshead:2018oaa}, integrals $\mathcal{I}_1$ and $\mathcal{I}_2$ were evaluated analytically and exactly.  Furthermore, it was found that in the $\mu \ll 1 \ll \xi$ range, adiabatic subtraction agreed with the result obtained by simply dropping off the divergent pieces.  Outside of this regime, the regularizating term generated by adiabatic regularization dominates the calculated contribution even outside of the UV limit, and furthermore, the value of the term depends on the order of adiabatic regularization.  Therefore, as in reference~\cite{Adshead:2018oaa}, we regularize by simply dropping the divergent pieces, noting that in the regime in which adiabatic regularization is well-behaved, these approaches agree.

We reproduce here the exact analytic results for $\mathcal{I}_1$ and $\mathcal{I}_2$ from reference~\cite{Adshead:2018oaa} along with their finite piece in the regime $\mu \ll 1 \ll \xi$.  The third integral, $\mathcal{I}_3$, can be evaluated with the same techniques.  The first integral is
\begin{align}
\mathcal{I}_1 &= \mu \left[ 
\frac{1}{2} \left(2 \Lambda^2 + \frac{1}{4} \left( \mu ^4-7 \mu ^2+12\right)-2 (\log (2 \Lambda )+\gamma_E ) \left(\mu ^2-8 \xi ^2+1\right)\right)
\right. \nonumber \\
& \qquad \left. + \frac{1}{4} \left(\mu ^2-2 \xi  (4 \xi +3 i)+1\right)
\left[H_{-i \left(2 \xi +\sqrt{\mu ^2+4 \xi ^2}\right)} \left(\sinh (4 \pi  \xi ) \text{csch}\left(2 \pi  \sqrt{\mu ^2+4 \xi ^2}\right)+1\right) \right. \right. \nonumber \\
&  \qquad\qquad \left. \left. +H_{i \left(\sqrt{\mu ^2+4 \xi ^2}-2 \xi \right)} \left(1-\sinh (4 \pi  \xi ) \text{csch}\left(2 \pi  \sqrt{\mu ^2+4 \xi ^2}\right)\right)\right] \right. \nonumber \\
&  \qquad\left. 
+\frac{1}{4} \left(\mu ^2-8 \xi ^2+6 i \xi +1\right) \left[H_{i \left(2 \xi +\sqrt{\mu ^2+4 \xi ^2}\right)} \left(\sinh (4 \pi  \xi ) \text{csch}\left(2 \pi  \sqrt{\mu ^2+4 \xi ^2}\right)+1\right)\right. \right. \nonumber \\
& \qquad \qquad \left. \left. +H_{-i \left(\sqrt{\mu ^2+4 \xi ^2}-2 \xi \right)} \left(1-\sinh (4 \pi  \xi ) \text{csch}\left(2 \pi  \sqrt{\mu ^2+4 \xi ^2}\right)\right)\right] \right. \nonumber \\
&  \qquad\left.
+6 \xi  \sqrt{\mu ^2+4 \xi ^2} \sinh (4 \pi  \xi ) \text{csch}\left(2 \pi  \sqrt{\mu ^2+4 \xi ^2}\right)
-\frac{\mu ^4}{8}
+\frac{11 \mu ^2}{8}-12 \xi ^2
\right], \nonumber \\
& \approx - 8\, \mu \,\xi^2\ln(\xi) .
\end{align}
The second integral reads
\begin{align}
\mathcal{I}_2
&= \Lambda ^4-\frac{\Lambda ^2 \mu ^2}{2}-\frac{7 \mu ^4}{16}+\mu ^2 \left(16 \xi ^2-\frac{19}{16}\right) -8 \xi ^4+\frac{11 \xi ^2}{2}
\nonumber \\
& \qquad +\frac{\xi}{4}   \left(-26 \mu ^2+16 \xi ^2-11\right) \sqrt{\mu ^2+4 \xi ^2} \sinh (4 \pi  \xi ) \text{csch}\left(2 \pi  \sqrt{\mu ^2+4 \xi ^2}\right) \nonumber \\
& \qquad + \frac{3}{16} \mu ^2 \left(\mu ^2-16 \xi ^2+1\right) \left[
4 (\log (2 \Lambda )+\gamma_E ) \right. \nonumber \\
& \qquad\qquad \left. + 
\left(H_{-i \left(\sqrt{\mu ^2+4 \xi ^2}-2 \xi \right)}
+H_{i \left(\sqrt{\mu ^2+4 \xi ^2}-2 \xi \right)}\right) 
\left(\sinh (4 \pi  \xi ) \text{csch}\left(2 \pi  \sqrt{\mu ^2+4 \xi ^2}\right)-1\right) \right. \nonumber \\
& \qquad\qquad \left. -
\left(H_{-i \left(2 \xi +\sqrt{\mu ^2+4 \xi ^2}\right)}
+H_{i \left(2 \xi +\sqrt{\mu ^2+4 \xi ^2}\right)}\right) 
\left(\sinh (4 \pi  \xi ) \text{csch}\left(2 \pi  \sqrt{\mu ^2+4 \xi ^2}\right)+1\right)\right], \nonumber \\
& \approx - 4 \pi\, \mu^2\, \xi^3 .
\end{align}
Finally, the third integral evaluates to
\begin{align}
\mathcal{I}_3 &= 
\frac{2}{3} \xi  \left(-6 \mu ^2+8 \xi ^2-1\right)+\frac{1}{3} \left(4 \mu ^2-8 \xi ^2+1\right) \sqrt{\mu ^2+4 \xi ^2} \sinh (4 \pi  \xi ) \text{csch}\left(2 \pi  \sqrt{\mu ^2+4 \xi ^2}\right)
\nonumber \\
& \qquad + \mu ^2 \xi  
\left[ 4 (\log (2 \Lambda )+\gamma_E ) \right. \nonumber \\
& \qquad \qquad \left.
+ \left(H_{-i \left(\sqrt{\mu ^2+4 \xi ^2}-2 \xi \right)}+H_{i \left(\sqrt{\mu ^2+4 \xi ^2}-2 \xi \right)}\right) \left(\sinh (4 \pi  \xi ) \text{csch}\left(2 \pi  \sqrt{\mu ^2+4 \xi ^2}\right)-1\right) \right. \nonumber \\
& \qquad \qquad \left. -\left(H_{-i \left(2 \xi +\sqrt{\mu ^2+4 \xi ^2}\right)}+H_{i \left(2 \xi +\sqrt{\mu ^2+4 \xi ^2}\right)}\right) \left(\sinh (4 \pi  \xi ) \text{csch}\left(2 \pi  \sqrt{\mu ^2+4 \xi ^2}\right)+1\right)\right], \nonumber \\
& \approx \frac{8}{3} \pi \, \mu ^2\, \xi ^2.
\end{align}
In these equations $H_n$ denotes the $n$-th harmonic number. From this we find final results for each loop,
\begin{align}
\left< \hat \gamma _{{\bf p}_1 }^{\lambda_1}(\tau) \hat \gamma _{{\bf p}_2}^{\lambda_2}(\tau) \right>_{1}
&=\dfrac{4\, H^4 \delta({\bf p}_1 + {\bf p}_2) \,\delta^{\lambda_1,\lambda_2}}{9 \pi\, M_{\rm Pl}^4\, p_1^3} \,\mu^2\,\xi^3
\,  \log(x) \,,
\nonumber  \\
\left< \hat \gamma _{{\bf p}_1 }^{\lambda_1}(\tau) \hat \gamma _{{\bf p}_2}^{\lambda_2}(\tau) \right>_{2}
&= 0 , \nonumber \\
\left< \hat \gamma _{{\bf p}_1 }^{\lambda_1}(\tau) \hat \gamma _{{\bf p}_2}^{\lambda_2}(\tau) \right>_{3}
&= \lambda_1\,\dfrac{ H^4  \delta({\bf p}_1 + {\bf p}_2 ) \delta^{\lambda_1, \lambda_2} 
}{3\,M_{\rm Pl}^4 \,p_1^3}\, \mu ^2\, \xi ^2\, ,
\nonumber \\
\!\!\!\!\!\!\!\!\!\!\!\!\!\!\!\!\!\!\!\!\!\!\!\!\!\!\!\!\!\!\!\!
\left< \hat \gamma _{{\bf p}_1 }^{\lambda_1}(\tau) \hat \gamma _{{\bf p}_2}^{\lambda_2}(\tau) \right>_{4}
&=\dfrac{ H^4 \delta^{\lambda_1,\lambda_2} \delta({\bf p}_1 + {\bf p}_2)}{9 \pi\, M_{\rm Pl}^4\, p_1^3}\,\mu ^2\, \xi ^3\,
\log(x) \,,
\nonumber \\
 \!\!\!\! \!\!\!\! \!\!\!\! \!\!\!\!
\left< \hat{\gamma} _{{\bf p}_1 }^{\lambda_1}(\tau) \hat{\gamma} _{{\bf p}_2}^{\lambda_2}(\tau) \right>_{5}
& = -
\dfrac{4\,H^4 \delta^{\lambda_1,\lambda_2} \delta({\bf p}_1 + {\bf p}_2 )}{9 \pi\, M_{\rm Pl}^4\, p_1^3}\, \mu ^2\, \xi ^3 \,\log(x) \,, \nonumber \\
\!\!\!\!\!\!\!\!\!\!\!\!\!\!\!\!\!\!\!\!\!\!\!\!\!\!\!\!\!\!\!\!
\left< \hat{\gamma} _{{\bf p}_1 }^{\lambda_1}(\tau) \hat{\gamma} _{{\bf p}_2}^{\lambda_2}(\tau) \right>_{6}
& =  -\dfrac{ 2\,H^4 \delta({\bf p}_1 + {\bf p}_2 ) \delta^{\lambda_1, \lambda_2}}{9 \pi\, M_{\rm Pl}^4\, p_1^3 }  
\, \mu ^2\, \xi ^3 \, ,
\nonumber \\
\!\!\!\! \!\!\!\! \!\!\!\! \!\!\!\!
\left< \hat \gamma _{{\bf p}_1 }^{\lambda_1}(\tau) \hat \gamma _{{\bf p}_2}^{\lambda_2}(\tau) \right>_{7}
& = - \dfrac{H^4 \delta^{\lambda_1,\lambda_2} \delta({\bf p}_1 + {\bf p}_2 )}{\pi\, M_{\rm Pl}^4 p_1^3 }\,\mu^2\, \xi^3\, \log(x) \,,
\label{eq:final_two_pt}
\end{align}
from which we conclude that the dominant contribution in the $\mu \ll 1 \ll \xi$ limit reads
\begin{align}
\sum_{\mathrm{vertices}} \left< \hat{\gamma} _{{\bf p}_1 }^{\lambda_1}(\tau) \hat{\gamma} _{{\bf p}_2}^{\lambda_2}(\tau) \right>_{\mathrm{quartic}}
&= -\frac{8\, H^4 \mu ^2 \xi ^3 \log (x)}{9 \pi\,  M_{\rm Pl}^4\, p_1^3} \delta^{\lambda_1,\lambda_2} \delta({\bf p}_1 + {\bf p}_2 )\,,
\label{eq:quartic_result}
\end{align}
which we also write as eq. (\ref{quartic-result}) of the main text. We conclude this appendix by noting that the parity violating vertex that originates from the portion of interaction Hamiltonian denoted by $H_{\rm {int},3}^{(4)}$.  The parity violation is evident because this term gives a contribution to the two-point function of the graviton which is proportional to the sign of the graviton helicity; however, the contribution is sub-dominant, scaling as ${\cal O}(\xi^2)$.

\section{Details of the cubic loop computation}
\label{ap:cubic}

In this appendix, we calculate the contribution from the one loop diagram with two cubic vertices shown in the right panel of figure \ref{fig:diagrams}, using the interaction term~(\ref{H3}) and the mode functions calculated in section~\ref{sec:mode_functions}.  We follow the approach, including the approximations, used in reference~\cite{Adshead:2018oaa}.  To start with, we drop the external momentum within the loop integral (${\bf k}_1 + {\bf p}_1 \rightarrow {\bf k}_1$); then after taking the $\tau \rightarrow 0$ we have
\begin{align}
 &\left< \hat{\gamma}^{\lambda_1}_{{\bf p}_1}(\tau)
\hat{\gamma}^{\lambda_2}_{{\bf p}_2}(\tau) \right>
=  \dfrac{H^4 \delta({\bf p}_1 + {\bf p}_2)\delta^{\lambda_1,\lambda_2}}{12\, M_{\rm Pl}^4\,  p_1^6} 
\int^\tau d\tau_1 [p_1\tau_1\, \cos (p_1\tau_1)-\sin (p_1\tau_1)]
 \nonumber \\
& 
\times\Big\{\int^{\tau_1} d\tau_2\, e^{i p_1\tau_2} (p_1\tau_2+i)
\sum_{r,s}
\int \dfrac{d^3k_1}{(2\pi)^3}
k_1^2
\left( 1 - \dfrac{sr}{5}  \right)  
\left[ rs {v}^r_{k_1}(\tau_1) {v}^s_{k_1}(\tau_1)
- {u}^r_{k_1}(\tau_1) {u}^s_{k_1}(\tau_1) \right]  \nonumber \\
&\times\left[
rs {v}^r_{k_1}(\tau_2)^* {v}^s_{k_1}(\tau_2)^*- {u}^r_{k_1}(\tau_2)^* {u}^s_{k_1}(\tau_2)^* \right] 
 + {\rm {h.c.}}\Big\}\,.
\end{align}

The evaluation of these integrals is complicated by the fact that the integrand is oscillating rapidly.  Therefore, we perform a Wick rotation, noting that $\tau_1 > \tau_2$ and, because of the nested integrals, both must be Wick rotated in the same direction.  This gives 
\begin{align}
\left< \hat{\gamma}^{\lambda_1}_{{\bf p}_1}(\tau)
\hat{\gamma}^{\lambda_2}_{{\bf p}_2}(\tau) \right>
&=-  \dfrac{H^4\, \delta({\bf p}_1 + {\bf p}_2)\delta^{\lambda_1,\lambda_2}}{60 \pi^2 M_{\rm Pl}^4  p_1^3} 
\int^{\infty}_{x} \dfrac{dx_1}{x_1^4} 
(x_1 \cosh(x_1)- \sinh(x_1) )
\int_{x_1}^{\infty} \dfrac{dx_2}{x_2} e^{-x_2} (1 + x_2)  \nonumber \\
& 
\times \sum_{r} 
\int dy_1 \, y_1^2\,\Bigg[ 3  \left(s^r(-iy_1) s^{-r}(-iy_1) + d^r(-iy_1) d^{-r}(-iy_1)\right) \nonumber \\
& \qquad  \times \left(s^r\(-i \frac{x_2 y_1}{x_1}\)^* s^{-r}\(-i \frac{x_2 y_1}{x_1}\)^* + d^r\(-i \frac{x_2 y_1}{x_1}\)^* d^{-r}\(-i \frac{x_2 y_1}{x_1}\)^*\right)
 \nonumber \\
& 
\qquad + 2 \left(s^r\(-i y_1\) d^r\(-i y_1\) + d^r\(-i y_1\) s^r\(-i y_1\)\right) \nonumber \\
& \qquad \times 
\left(s^r\(-i \frac{x_2 y_1}{x_1}\)^* d^r\(-i \frac{x_2 y_1}{x_1}\)^* + d^r\(-i \frac{x_2 y_1}{x_1}\)^* s^r\(-i \frac{x_2 y_1}{x_1}\)^*\right)
 +{\rm h.c.} \Bigg],
\end{align}
where $x_1 = - p_1 \tau_1$, $x_2 = - p_1 \tau_2$, and $y_1 = - k_1\tau_1$, and functions such as $s^r(-iy_1)^*$ should be interpreted as first complex conjugating $s^r(x)$ and then substituting $x = - i y_1$.

Next we introduce polar coordinates in the $x_1-x_2$ plane,
\begin{align}
x_1 = \rho \cos(\alpha), \qquad
x_2 = \rho \sin(\alpha).
\end{align}

\begin{figure}
\begin{center}
\includegraphics[scale=1]{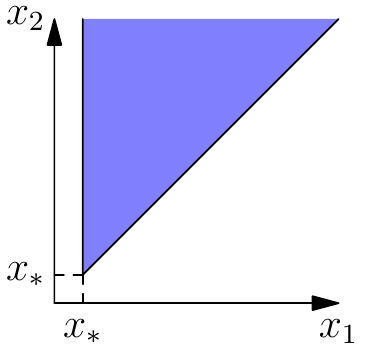} \hspace*{1cm}
\includegraphics[scale=1]{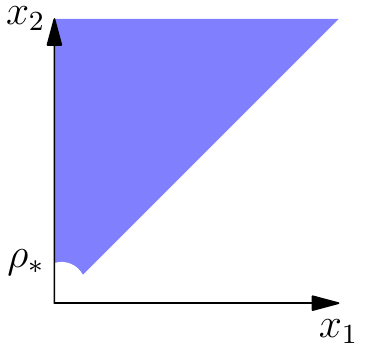}
\caption{\label{fig:int_cont} Left: The actual region of integration in the $x_1 -x_2$ plane.  Right: The region of integration used.  At fixed $y_1$, the integrand is exponentially suppressed as the polar angle $\alpha \rightarrow \pi \slash 2$, and so this replacement introduces at most an $\mathcal{O}(1)$ uncertainty.  }
\end{center}
\end{figure}

The true region of integration is shown on the left side of figure~\ref{fig:int_cont}; however, we approximate the integral by using the region of integration shown on the right side.  We show below that at fixed $y_1$, the integral is exponentially suppressed as $\alpha \rightarrow \pi \slash 2$, and therefore we expect this to contribute at most a $\mathcal{O}(1)$ factor.

After further substituting $\beta = \tan(\alpha)$, we can perform the  integral in $d \rho$ to find
\begin{align}
 \left< \hat{\gamma}^{\lambda_1}_{{\bf p}_1}(\tau)
\hat{\gamma}^{\lambda_2}_{{\bf p}_2}(\tau) \right>
& =  \dfrac{H^4\, \delta({\bf p}_1 + {\bf p}_2)\delta^{\lambda_1,\lambda_2}}{540 \pi^2 M_{\rm Pl}^4  p_1^3} \nonumber \\
& \times
\int_{1}^{\infty} \frac{d\beta}{\beta} \,  
\left(3 \beta ^3 \coth ^{-1}\beta -3 \beta ^2-3 \coth ^{-1}\left(\beta ^2\right)+3 \log \rho_*+3 \gamma_E -4\right)
 \nonumber \\
& 
\times \sum_{r} 
\int dy_1 \, y_1^2\Bigg[ 3  \left(s^r(-iy_1) s^{-r}(-iy_1) + d^r(-iy_1) d^{-r}(-iy_1)\right)  \nonumber \\
& \times \left(s^r\(-i \frac{x_2 y_1}{x_1}\)^* s^{-r}\(-i \frac{x_2 y_1}{x_1}\)^* + d^r\(-i \frac{x_2 y_1}{x_1}\)^* d^{-r}\(-i \frac{x_2 y_1}{x_1}\)^*\right)
  \nonumber \\
&  + 2 \left(s^r(-i y_1) d^r(-i y_1) + d^r(-i y_1) s^r(-i y_1)\right)\nonumber \\
&\times
\left(s^r\(-i \frac{x_2 y_1}{x_1}\)^* d^r\(-i \frac{x_2 y_1}{x_1}\)^* + d^r\(-i \frac{x_2 y_1}{x_1}\)^* s^r\(-i \frac{x_2 y_1}{x_1}\)^*\right)
 +{\rm h.c.} \Bigg],
\end{align}
in the $\rho_* \rightarrow 0$ limit.  This is dominated by the $\log(\rho_*)$ piece, which is regulated by the finite number of e-foldings between when the mode leaves the horizon and the end of inflation; the same behavior was also observed in the quartic loops in appendix~\ref{ap:quartic} above. Physically, this logarithmic behavior is a consequence of the fact that the nonzero fermion energy density continues to source graviton fluctuations even once it is outside the horizon.

Finally, we approximate the Whittaker functions as in reference~\cite{Adshead:2018oaa}.  Along the positive axis we use (defining $M\equiv\sqrt{\mu^2+4\,\xi^2}$)
\begin{align}\label{eq:Positive_Axis}
W_{1/2 - 2 i r\xi, i M}(2x>0)
& \approx \dfrac{(2x)^{-ir M+1 \slash 2} e^{-x}}{\Gamma(i r M + 2 i r \xi)} \Gamma(2 i r M) ,
\nonumber \\
W_{-1/2 - 2 i r \xi, i M}(2x>0)
& \approx \dfrac{(2x)^{-ir M+1 \slash 2} e^{-x}}{\Gamma(i r M + 2 i r \xi+1)} \Gamma(2 i r M),
\end{align}
and we note that the Whittaker functions are even in their second index.  Along the negative axis, we use 
\begin{align}
W_{1/2+2 i r\xi ,-i M}(2 x<0)
&\approx \mathcal{A}_{1r} e^{-2 \pi \xi} e^{-x} x^{\frac{1}{2}+ir M}
+ \mathcal{B}_{1r} e^{-2 \pi  \xi } e^{x} (-x)^{\frac{1}{2}-ir M}, \nonumber \\
W_{-1/2+ 2 i r\xi ,-i M}(2 x<0)
&\approx \mathcal{A}_{2r} e^{-2 \pi \xi} e^{-x} x^{\frac{1}{2}+ir M}
+ \mathcal{B}_{2r} e^{-2 \pi \xi} e^x (-x)^{\frac{1}{2}-ir M},
\label{eq:Negative_Axis}
\end{align}
where
\begin{alignat}{2}
\mathcal{A}_{1r} &= -\frac{2^{\frac{1}{2}+ir M} e^{-2 \pi  \xi} \Gamma \left(-2 ir M\right)}{\Gamma \left(-ir \left(2 \xi +M\right)\right)}, \qquad &
\mathcal{B}_{1r} &= \frac{ir  2^{\frac{3}{2}-ir M}  \Gamma \left(2 ir M\right) \sinh \left(\pi  \left(M+2 \xi \right)\right)}{\Gamma \left(ir \left( M-2 \xi\right)\right)}, \nonumber \\
\mathcal{A}_{2r} &= -\frac{2^{\frac{1}{2}+ir M} e^{-2 \pi  \xi} \Gamma \left(-2 ir M\right) }{\Gamma \left(-2 ir \xi -ir M+1\right)}, \qquad &
\mathcal{B}_{2r} &= \frac{ir  2^{\frac{3}{2}-ir M} \Gamma \left(2 ir M\right) \sinh \left(\pi  \left(M+2 \xi \right)\right)}{\Gamma \left(-2 ir \xi +irM+1\right)}.
\end{alignat}

With these approximations, we find
\begin{align}\label{eq:appr_cubic}
& \left< \hat{\gamma}^{\lambda_1}_{{\bf p}_1}(\tau)
\hat{\gamma}^{\lambda_2}_{{\bf p}_2}(\tau) \right>
=  \dfrac{H^4\, \delta({\bf p}_1 + {\bf p}_2)\delta^{\lambda_1,\lambda_2}}{180 \pi^2 M_{\rm Pl}^4  p_1^3}  \log(\rho_*)
\int_{1}^{\infty} \dfrac{d\beta}{\beta} 
\sum_{r}  \int dy_1 \, y_1^2\, e^{-2 \beta  y_1}\, e^{-4 \pi \xi}
 \nonumber \\
& \qquad
\times\Big\{\Big[
 3 \mathcal{C}_{1,r}\, \beta\,  y_1^2  
\left[
\left( - \mathcal{A}_{1,r} \mathcal{A}_{1,-r} + \mu^2 \mathcal{A}_{2,r} \mathcal{A}_{2,-r} \right) e^{2 y_1}
+ \left( \mathcal{B}_{1,r} \mathcal{B}_{1,-r} - \mu^2 \mathcal{B}_{2,r} \mathcal{B}_{2,-r} \right) e^{-2 y_1}\right. \nonumber \\
& \qquad\qquad\left. + i \left( \mathcal{A}_{1,r} \mathcal{B}_{1,-r} - \mu^2 \mathcal{A}_{2,r} \mathcal{B}_{2,-r} \right)
(-y_1^2)^{ir M} 
+ i \left( \mathcal{B}_{1,r} \mathcal{A}_{1,-r} - \mu^2 \mathcal{B}_{2,r} \mathcal{A}_{2,-r} \right)
(-y_1^2)^{-ir M}
\right]   \nonumber \\
& \qquad\qquad+ 8 \,\mu^2 \,e^{- 4 \pi r \xi} \, \mathcal{C}_{2,r} \, y_1^2\, \beta ^{1-2 i M r}
\left[- \mathcal{A}_{1,r} \,\mathcal{A}_{2,r} \,e^{-2 \pi  M r +2 y_1 }
  + 
\mathcal{B}_{1,r}\, \mathcal{B}_{2,r} 
 y_1^{-4 i M r} e^{-2 y_1}  \right.  \nonumber \\
 & \qquad\qquad\qquad\qquad\qquad\qquad\qquad\quad\qquad\left.  +i\, y_1^{-2 i M r} \,
 e^{- M \pi  r} \left( \mathcal{A}_{1,r}\, \mathcal{B}_{2,r} 
  + \mathcal{A}_{2,r}\, \mathcal{B}_{1,r} 
\right)  \right]
\Big] 
+{\rm  h.c.}
\Big\},
\end{align}
where we have defined
\begin{align}
\mathcal{C}_{1,r} &= \dfrac{4\,\xi}{M}\, \text{csch}(2 \pi  M r) \sinh (\pi  r (M+2 \xi )), \nonumber \\
\mathcal{C}_{2,r} &= \frac{\Gamma (2 i M r)^2  2^{1-2 i M r}}{\Gamma (i M r+2 i \xi  r) \Gamma (i M r+2 i \xi  r+1)}.
\end{align}

Eq.~(\ref{eq:appr_cubic}) shows that, as claimed, at fixed $y_1$, this contribution to the graviton two-point function  is exponentially suppressed as $\beta \rightarrow \infty$ ($\alpha \rightarrow \pi \slash 2$).  The $dy_1$ integral may now be performed analytically, leaving the $d\beta$ integral.  We separate this into parts involving $\mathcal{A}$ only, $\mathcal{B}$ only, and mixed.  From the form~\eqref{eq:Negative_Axis}, we recognize that the $\mathcal{A}$ terms correspond to the vacuum part of the modes, which are nonzero and positive frequency as $p \rightarrow \infty$.  We subtract this part by hand; we also note that it is only the purely $\mathcal{A}$ part which has a divergent piece.  We regularize this by writing the lower limit of the $d\beta$ integral as $1 + \epsilon$ and taking the limit as $\epsilon \rightarrow 0$. 

Taking the large $\xi$ limit, the purely $\mathcal{A}$ piece which we renormalize away is 
\begin{align}
\left< \hat{\gamma}^{\lambda_1}_{{\bf p}_1}(\tau)
\hat{\gamma}^{\lambda_2}_{{\bf p}_2}(\tau) \right>_{\mathcal{A}\mathcal{A}}
& = - \dfrac{H^4\, \delta({\bf p}_1 + {\bf p}_2)\delta^{\lambda_1,\lambda_2}}{240 \pi^2 M_{\rm Pl}^4  p_1^3}  \log(\rho_*) \mu^2
\left\lbrace -\frac{8 }{\epsilon ^2}+\frac{16 }{\epsilon } \right. \nonumber \\
& \qquad \left. -\frac{16}{9}   \xi ^2 \left(6 H_{4 i \xi +3}+6 H_{3-4 i \xi }+12 \log (\epsilon )-25\right)
\right\rbrace,
\end{align}
where we have dropped a term that, in the large $\xi$ limit, is proportional to $\xi ^{-2} \epsilon ^{-4}$, and where $H_n$ is the $n$-th harmonic number.  The mixed term is
\begin{align}\label{eq:mixedterm}
\left< \hat{\gamma}^{\lambda_1}_{{\bf p}_1}(\tau)
\hat{\gamma}^{\lambda_2}_{{\bf p}_2}(\tau) \right>_{\mathcal{A}\mathcal{B}}
& =  \dfrac{16\,H^4\, \delta({\bf p}_1 + {\bf p}_2)\delta^{\lambda_1,\lambda_2}}{45\pi\, M_{\rm Pl}^4\,  p_1^3}\,\mu ^2\,\xi ^3  \log(\rho_*)\,,
\end{align}
while the purely $\mathcal{B}$ contribution is
\begin{align}
\left< \hat{\gamma}^{\lambda_1}_{{\bf p}_1}(\tau)
\hat{\gamma}^{\lambda_2}_{{\bf p}_2}(\tau) \right>_{\mathcal{B} \mathcal{B}}
& =  \dfrac{H^4\, \delta({\bf p}_1 + {\bf p}_2)\delta^{\lambda_1,\lambda_2}}{180 \pi^2 \,M_{\rm Pl}^4\,  p_1^3}  \log(\rho_*)
\left\lbrace
- \frac{9}{8} \sinh \left(\frac{\pi  \mu ^2}{4 \xi }\right) \right. \nonumber \\
& \qquad
\left. 
+ (1-i)\, \pi ^{3/2}\, \mu ^4 \,2^{\frac{17}{2}+8 i \xi }\, e^{4 \pi  \xi }\, \xi ^{3/2}\, B_{-1}(4 i \xi +4,-8 i \xi -4) \right. \nonumber \\
& \qquad \left.
+ (1+i)\, \pi ^{3/2}\, \mu ^4 \,2^{\frac{17}{2}-8 i \xi }\, e^{4 \pi  \xi }\, \xi ^{3/2}\, B_{-1}^*(4 i \xi +4,-8 i \xi -4)
\right\rbrace,
\end{align}
where $B_{n}(x,\,y)$ denotes the incomplete beta function. This term has a piece which scales as $\mu^2 \slash \xi$ and a piece which scales approximately as $\mu^4 \xi$.  In the regime $\mu \lesssim 1$ and $\xi \gg 1$ either can dominate, depending on how small $\mu$ is.  Regardless, the mixed term~(\ref{eq:mixedterm}) dominates, and the leading contribution from the cubic diagram is
\begin{align}
\left< \hat{\gamma}^{\lambda_1}_{{\bf p}_1}(\tau)
\hat{\gamma}^{\lambda_2}_{{\bf p}_2}(\tau) \right>_{\mathcal{A}\mathcal{B}}
& \sim  \mathcal{O}(0.1) \times \dfrac{H^4 \delta({\bf p}_1 + {\bf p}_2)\delta^{\lambda_1,\lambda_2}}{M_{P}^4\,  p_1^3}  \mu ^2 \,\xi ^3\, \log(\rho_*),
\end{align}
where we have noted an overall uncertainty due to the approximations used.  We also note that $\rho_* \sim - p_1 \tau$ from the change to polar coordinates.  This expression is reported as eq. (\ref{cubic-result}) of the main text. This is the same order of magnitude as the quartic result in eq.~\eqref{eq:quartic_result}, although with the opposite sign.

\bibliography{APPRS19}
\bibliographystyle{JHEP}

\end{document}